\documentclass[journal=pasa,manuscript=research-article,year=2024,volume=41,]{cup-journal}

\usepackage{amsmath}
\usepackage{microtype}
\usepackage{booktabs}
\usepackage{amssymb}

\newcommand{\XMM}{XMM-Newton}
\newcommand{\psr}{PSR~J0837$-$2454}
\newcommand{\Msun}{M_\odot}
\newcommand{\NH}{N_{\rm H}}
\newcommand{\Tinfty}{T^\infty}
\newcommand{\Rinfty}{R^\infty}
\newcommand{\Linfty}{L^\infty}

\newcommand{\Teff}{T_{\rm eff}^\infty}
\newcommand{\Tinftyb}{T^\infty_2}
\newcommand{\Rinftyb}{R^\infty_2}
\newcommand{\Tc}{T_{\rm c}}

\title{First detection of X-ray pulsations and spectrum of the high Galactic latitude pulsar \psr\ and direct Urca cooling implications}

\author{Wynn C. G. Ho}
\affiliation{Department of Physics and Astronomy, Haverford College, 370 Lancaster Avenue, Haverford, PA 19041, USA}
\email[W.C.G. Ho]{who@haverford.edu}

\author{Nihan Pol}
\affiliation{Department of Physics, Oregon State University, Corvallis, OR 97331, USA;
Center for Gravitation, Cosmology and Astrophysics, Department of Physics, University of Wisconsin-Milwaukee, P.O. Box 413, Milwaukee, WI 53201, USA}

\author{Adam T. Deller}
\affiliation{Centre for Astrophysics and Supercomputing, Swinburne University of Technology, John Street, Hawthorn, VIC, 3122, Australia}

\author{Werner Becker}
\affiliation{Max-Planck-Institut f\"{u}r extraterrestrische Physik, Gie{\ss}enbachstra{\ss}e 1, 85748 Garching, Germany;
Max-Planck-Institut f\"{u}r Radioastronomie, Auf dem H\"{u}gel 69, 53121 Bonn, Germany}

\author{Sarah Burke-Spolaor}
\affiliation{Department of Physics and Astronomy, West Virginia University, P.O. Box 6315, Morgantown, WV 26506, USA;
Center for Gravitational Waves and Cosmology, West Virginia University, Chestnut Ridge Research Building, Morgantown, WV 26505, USA}


\begin{document}

\begin{abstract}
\psr\ is a young 629~ms radio pulsar whose uncertain distance has important
implications.
A large distance would place the pulsar far out of the Galactic plane and
suggest it is the result of a runaway star, while a short distance would
mean the pulsar is extraordinarily cold.
Here we present further radio observations and the first deep X-ray observation of \psr.
Data from the Parkes Murriyang telescope show flux variations over short
and long timescales and also yield an updated timing model,
while the position and proper motion (and, less strongly, parallax) of the pulsar are constrained by a number of low-significance detections with the Very Long Baseline Array.
\XMM\ data enable detection of X-ray pulsations for the first time
from this pulsar and yield a spectrum that is thermal and blackbody-like,
with a cool blackbody temperature $\approx70\mbox{ eV}$ or atmosphere
temperature $\approx50\mbox{ eV}$, as well as a small hotspot.
The spectrum also indicates the pulsar is at a small distance of
$\lesssim 1\mbox{ kpc}$, which is compatible with the marginal VLBA parallax constraint that favours a distance of $\gtrsim330$~pc.
The low implied luminosity
($\sim7.6\times10^{31}\mbox{ erg s$^{-1}$}$ at 0.9~kpc)
suggests \psr\ has a mass high enough that fast neutrino emission from
direct Urca reactions operates in this young star and points to a nuclear
equation of state that allows for direct Urca reactions at the highest
densities present in neutron star cores.
\end{abstract}

\section{Introduction} \label{sec:intro}

Although radio emission was detected earlier \citep{Burkespolaoretal11},
discovery of the spin period of \psr\ and measurements of the pulsar's
timing parameters were only made recently \citep{Poletal21}.
These measurements yielded a pulsar period $P=629\mbox{ ms}$ and
spin period time derivative $\dot{P}=3.49\times10^{-13}\mbox{ s s$^{-1}$}$,
which imply a characteristic spin-down age $\tau_{\rm c}=28600\mbox{ yr}$,
magnetic field $B=1.5\times10^{13}\mbox{ G}$,
and spin-down luminosity $\dot{E}=5.5\times10^{34}\mbox{ erg s$^{-1}$}$.
These parameters suggest \psr\ is a young, strongly magnetic,
mildly energetic pulsar but not particularly unique among the
general pulsar population,
although it is apparently also a rotating radio transient (RRAT;
see RRATalog\footnote{http://astro.phys.wvu.edu/rratalog/} and
\citealt{Abhisheketal22}).
What makes \psr\ special is its three-dimensional spatial position
and the implications of this position.

\psr\ is located at
R.A.~$=08^{\rm h}37^{\rm m}57.73^{\rm s}\pm0.06^{\rm s}$,
decl.~$=-24^\circ54'30''\pm1''$ (J2000), which leads to a Galactic
longitude and latitude of $247.58^\circ$ and $9.77^\circ$, respectively
\citep{Poletal21}.
The high latitude is particularly noteworthy since this means that
\psr\ is possibly located far above the Galactic plane.
The radio observations of \citet{Poletal21} gave a dispersion measure (DM) of
$143.1\pm0.1\mbox{ pc cm$^{-3}$}$ from scattering by interstellar
electrons, which implies a pulsar distance of 6.3~kpc using the
NE2001 model of the Galactic electron density \citep{CordesLazio02}
or $>25\mbox{ kpc}$ using the YMW16 model \citep{Yaoetal17}.
At the former distance, the pulsar's height above the plane is 1.1~kpc,
while the latter distance gives a height $\gg 1\mbox{ kpc}$.
Thus if the large distance inferred from the radio measurements
is correct, then \psr\ is far out of the Galactic plane and is very likely to
be one of the only known pulsars to be born from a runaway O/B star
(see \cite{Poletal21}, for more details),
and its characteristics would provide insights into the pulsar's
origin and connection to other neutron stars.

On the other hand,
pulsar distances inferred using radio dispersion measures can
occasionally be discrepant from distances measured via other techniques
(see, e.g., \citealt{Delleretal19}).
In the case of \psr, radio imaging shows diffuse emission around
\psr, which could imply a much lower distance if associated with
the pulsar.
In particular,
GLEAM radio data show a $\sim1.5^\circ$ diameter circular structure
surrounding \psr\ that could be a supernova remnant \citep{Hurleywalkeretal17}.
If this structure is indeed a supernova remnant and is the remnant
left over from the birth of \psr, then the distance is estimated
to be 0.9~kpc \citep{Poletal21}.
Alternatively, H$\alpha$ observations from SHASSA \citep{Gaustadetal01}
show enhanced diffuse emission around \psr\ (but not coincident with
the potential supernova remnant), which could contribute to the
aforementioned dispersion measure and could imply an even lower
distance of 0.2~kpc \citep{Poletal21}.
At these two short distance possibilities, the pulsar's height above
the Galactic plane is either 151 or 34~pc and thus not particularly
unusual.
However, what would be extraordinary in this case is the pulsar's low X-ray
luminosity, which would make \psr\ one of the coolest young neutron stars known,
and its characteristics would provide crucial constraints on neutron star
cooling theory and nuclear physics properties.

A short 1.5~ks Swift XRT observation was taken on 2021 April 28
(ObsID 00014291001) and detected \psr\ at a 0.2--10~keV count rate of
0.0013~counts~s$^{-1}$ and an estimated 0.3--10~keV flux of
$\sim 7\times10^{-14}\mbox{ erg s$^{-1}$ cm$^{-2}$}$.
\psr\ was also detected by eROSITA in eRASS:4 but only at 0.2--0.7~keV,
indicating it is a very soft X-ray source \citep{MayerBecker24}.
The Swift flux means that the X-ray luminosity of \psr\ is
$L_{\rm X}\sim 7\times10^{30}\mbox{ erg s$^{-1}$}$ at a distance of 0.9~kpc
(note this luminosity estimate does not account for X-ray absorption or
spectral characteristics).
Young neutron stars of age $\sim 3\times 10^4\mbox{ yr}$
have thermal X-ray luminosities $\sim 10^{32}-10^{33}\mbox{ erg s$^{-1}$}$
(see, e.g., \citealt{Potekhinetal20}),
and young pulsars often have a non-thermal magnetosphere or pulsar wind
component to their X-ray luminosities (see, e.g., \citealt{Becker09}).
Cool neutron stars like \psr\ (and PSR~J0007+7303, PSR~B1727$-$47, and
PSR~B2334+61) are crucial sources because their low luminosities point to
cooling that invoke fast direct Urca neutrino emission processes,
which is only allowed for high core proton fractions and only predicted
by some nuclear equations of state
(\citealt{Lattimeretal91,PageApplegate92};
see, e.g., \citealt{Potekhinetal15}, for review).
Thus the few sources like \psr\ are critical astronomical objects for
probing fundamental physics at supranuclear densities.

Here we report the analyses of new radio observations using the Parkes Murriyang telescope and Very Long Baseline Array (VLBA) and the first long X-ray observations of \psr{}.
In Section~\ref{sec:obs}, we discuss the observational data and our analysis procedure, while in Section~\ref{sec:results}, we describe results from these observing campaigns.
In Section~\ref{sec:discuss}, we discuss the implications of our results for the distance to \psr{} and the low measured luminosity and surface temperature in the context of how neutron stars cool over time.
In Section~\ref{sec:summary}, we summarize our findings.

\section{Observations and data analysis} \label{sec:obs}

\subsection{Radio} \label{sec:radiodata}

The biggest uncertainty around \psr\ is that the dispersion measure derived distance ($d>6$~kpc) places this pulsar at the edge of the Galaxy and at an anomalous $z$-height ($>1$~kpc), while a distance estimated from diffuse emission in optical and radio give a much lower value ($d<1$~kpc) \citep{Poletal21}.
To conclusively determine the distance, we conducted observations of \psr\ using the VLBA to measure an astrometric parallax. We coupled these VLBA observations with observations using the ultra-wide Low (UWL) receiver \citep{UWL} on the Parkes Murriyang radio telescope, in order to produce a timing solution for gating the VLBA data and to study the spectro-temporal properties of the pulsar with the wide bandwidth and sensitivity of the UWL . Here we describe our observations using both instruments.

\subsubsection{Parkes Murriyang observations}

We obtained UWL observations of \psr\ from 2021 October to 2022 March and 2022 October to 2023 March for a total of 18 observing epochs. These observation windows were chosen to overlap with our VLBA observations so that we could mitigate the large red noise of this pulsar \citep{Poletal21} and produce an accurate timing solution for VLBA gating observations. The first six epochs in the first observing window had a total integration time of 1~hour, while the other epochs in both windows had a total integration time of 30~minutes. The observations used the {\sc medusa} backend, with observations performed in the ``fold'' mode, 30~s sub-integration time, 1~MHz resolution across the full 704--4032~MHz band, 1024 phase bins across one rotation, and full Stokes polarization information. In addition to the pulsar, each observation also monitored the local noise diode for flux and polarization calibration.

We processed the data using {\sc psrchive} and {\sc pypulse}. We identified and zapped radio frequency interference (RFI) using {\sc clfd} \citep{CLFD}, which is a smart RFI removal algorithm. Any sources of RFI missed by this algorithm were manually zapped using {\sc pazi}. We used separate, archival observations of PKS~B0407$-$658 for flux calibration, and long-track observations of PSR~J0437$-$4715 were used as the input to the polarization calibration {\sc pcm} routine in {\sc psrchive} for polarization calibration. These files were used with {\sc pac -cT} to calibrate the flux and polarization for each epoch.

Following calibration, we measured the rotation measure (RM) at each epoch using the brute force approach in {\sc rmfit}, where we searched for RMs between $-$3000~rad~m$^{-2}$ to 3000~rad~m$^{-2}$ with a step size of 6~rad~m$^{-2}$. We use {\sc ionFR} \citep{ionFR} to calculate the ionospheric Faraday rotation contribution to the measured total RM. We use the International GNSS Service global ionospheric total electron content maps \citep{igsg_1, igsg_2} and the International Geomagnetic Reference Field \citep{igrf_1} as inputs to {\sc ionFR}. We obtain ionospheric RMs in the range of $-$3.15~rad~m$^{-2}$ to $-$0.65~rad~m$^{-2}$ (negative values in the Southern hemisphere due to the orientation of Earth's geomagnetic field), which are used to obtain the non-terrestrial RM from $\rm RM_{\rm ISM} = RM_{\rm tot} - RM_{\rm ion}$.
The best-fit RM value was used to correct the observation for Faraday rotation with {\sc pam -R}.

\subsubsection{Very Long Baseline Array observations} \label{sec:vlbiobs}

We observed \psr\ using VLBA a total of twelve times between 2022 October and 2024 May (project code BD256). Observations were made using $8\times32$~MHz sub-bands placed in the range 1392--1744~MHz and avoiding known regions of radio frequency interference. Each observation was 1.5~hours in duration, spending around 60\% of the time on the target field and interleaving with 50~s scans on a phase reference source (ICRF~J084656.6$-$260750) every $\sim3$~minutes. The bright source IERS~B0742+103 was observed at the end of each observation to calibrate the instrumental bandpass.

Data were correlated using the DiFX software correlator and employing an ephemeris derived from Parkes monitoring to provide pulse-phase-resolved gating to improve the image signal-to-noise ratio on the pulsar field \citep{Delleretal11}. For most observations, two such correlation passes were run -- one producing a single ``matched filter'' output in which the pulse was weighted by the expected profile and summed in phase, and another in which 20 equally spaced pulsar ``bins'' were produced, each spanning 5\% of the pulse phase. In addition to these pulsar datasets, we produced a number of other visibility datasets from scans on the pulsar. First, a dataset centred at the pulsar position without any pulse-phase-resolved gating was used (the ``ungated'' pulsar dataset). Additionally, three other correlator passes were undertaken at the position of other compact radio sources visible within the VLBA primary beam, in order to provide information that could be used for calibration refinement. In a preliminary observation of the \psr\ field taken prior to the first astrometric observation, we imaged every known radio source within $\sim$25~arcminutes of \psr\ and identified three sources with flux densities ranging from 7--20~mJy that contained compact source components detectable by VLBA. Correlations centred at each of these positions provided three ``in-beam calibrator'' datasets.

Data reduction made use of the \verb+psrvlbreduce+ pipeline from \citet{Dingetal23} and employed standard phase and amplitude calibration techniques as described therein. After calibration, a $2''\times2''$ area centred on the nominal pulsar position was imaged, using both the gated and the ungated datasets (to guard against the eventuality that the pulsar ephemeris was insufficiently well modelled at the time of any given observation). Naturally weighted images, with a typical angular resolution of $5\times25$~milliarcseconds, were made at all epochs, with a typical image rms of $\sim60$~$\mu$Jy/beam in the ungated image. \psr\ was detected in 5 of 12 observations with a significance ranging from 7.5$\sigma$ to 13$\sigma$ in one bin of the binned datasets. The pulsar was not detected in the first three observations - as can be seen in Figure~\ref{fig:flux_var}, this corresponds to a time at which the flux density is at a minimum value. In the 4th and 5th observation, no ``binned'' datasets were produced, and the pulsar was not detected in the matched-filter or ungated datasets. At the time of submission, the final two observations had not yet been correlated and released.
Where detected, the cleaned VLBA images of \psr\ show a source size which is inconsistent with a point-source, demonstrating that the pulsar is significantly affected by angular broadening due to multi-path propagation in the turbulent ionised interstellar medium. Because the point spread function of the interferometer is elongated at these low declinations (a typical beam size was $5\times20$~milliarcseconds), useful constraints on the intrinsic source size are best obtained under the assumption of a circularly symmetric scatter--broadened source, although it is also possible that the scatter--broadening is mildly asymmetric. These returned a source size in the range 8--10~millisarcseconds. The extragalactic calibrator sources that were observed in--beam contemporaneously with the pulsar showed a slightly larger degree of angular broadening, with sizes ranging from 12--20~milliarcseconds, although this may also partly reflect intrinsic source structure.

\subsection{X-ray}

\begin{figure*}[hbt!]
\centering
\includegraphics[width=0.3\linewidth]{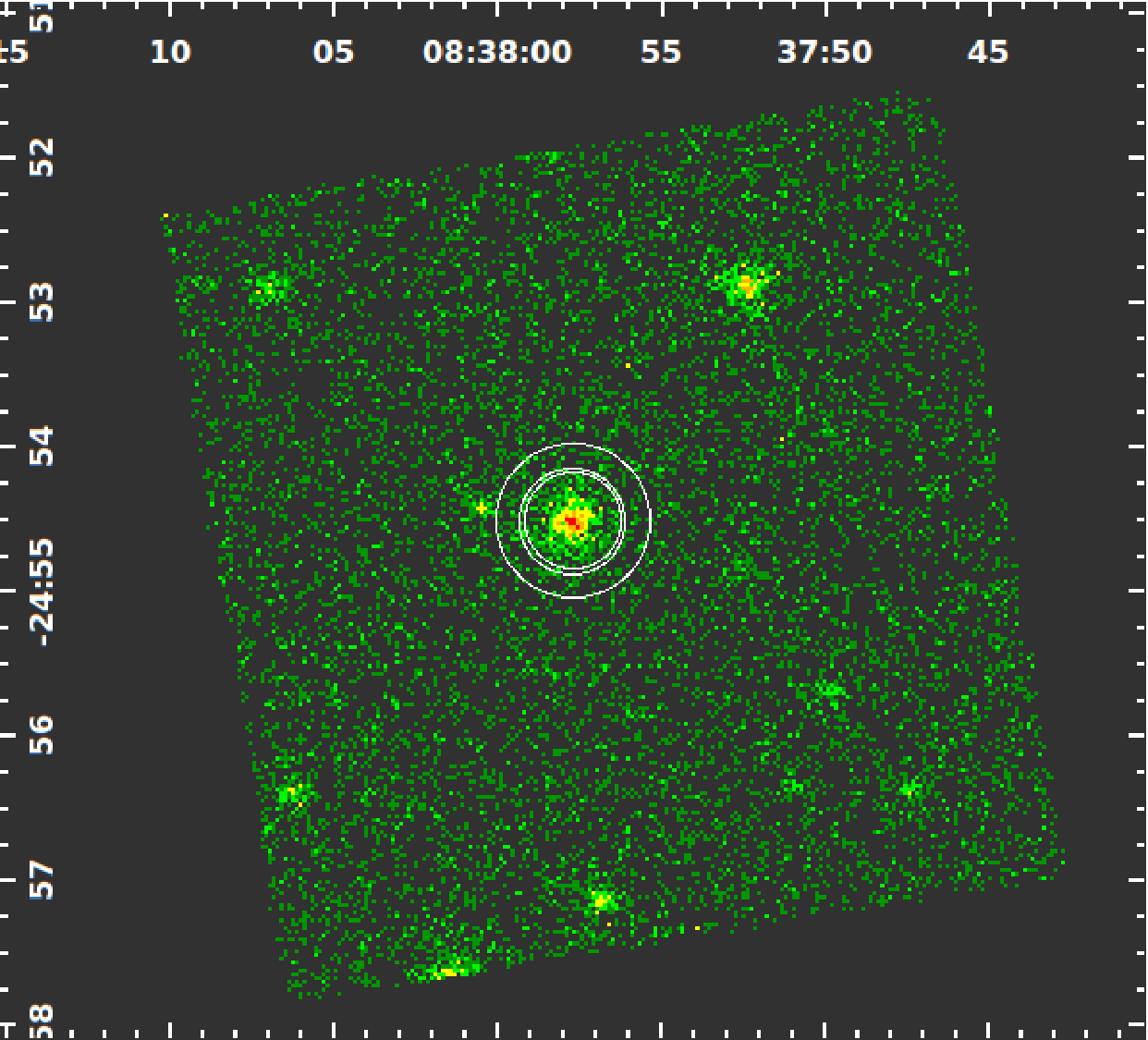}
\hspace{1em}
\includegraphics[width=0.3\linewidth]{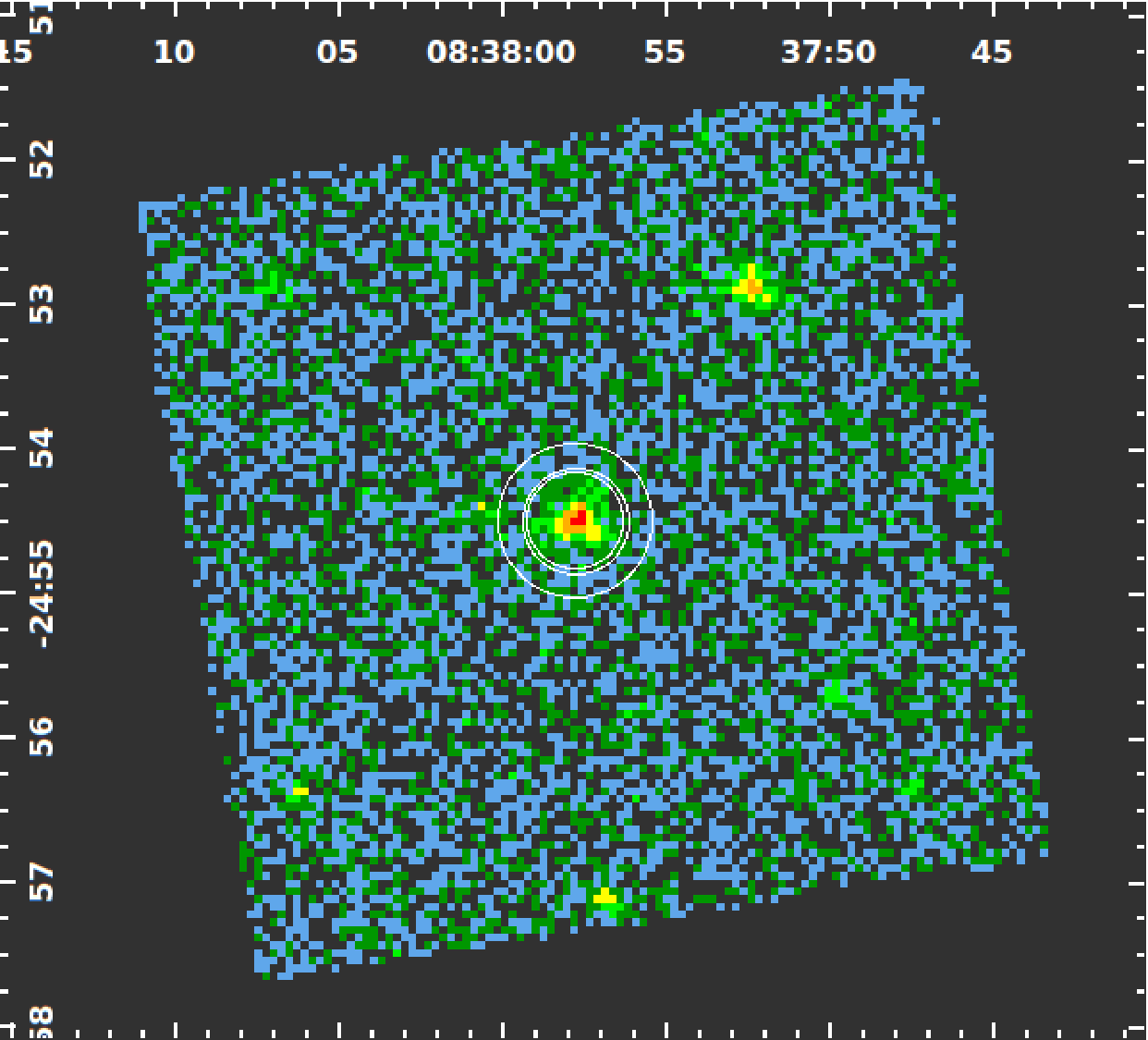}
\hspace{1em}
\includegraphics[width=0.3\linewidth]{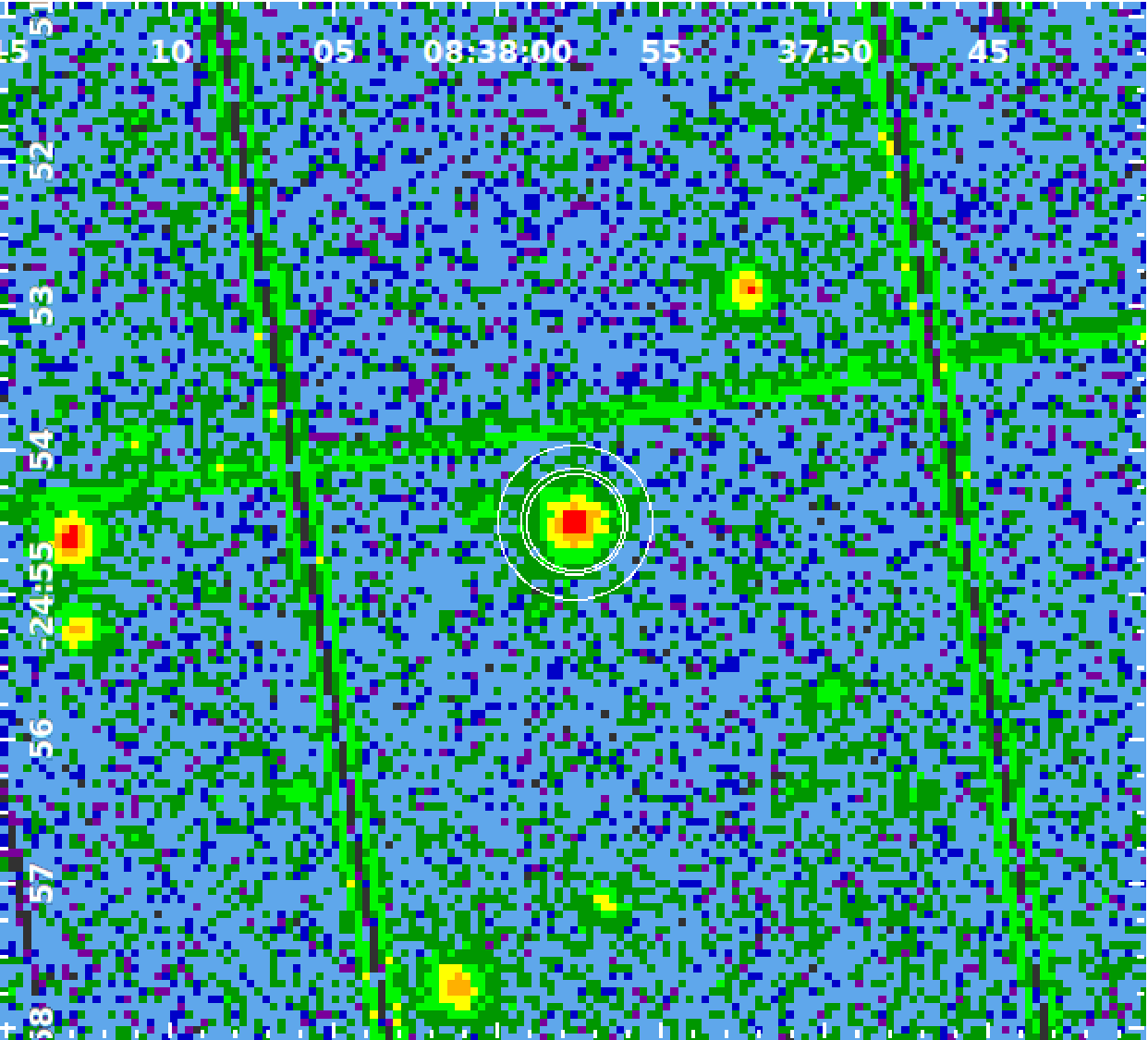}
\caption{X-ray field of \psr\ from \XMM\ MOS1 (left), MOS2 (middle),
and pn (right) in large window mode.
Source counts are extracted from a $20''$ radius circle, while background
counts are extracted from an annulus with inner and outer radii of $22''$
and $32''$, respectively.}
\label{fig:fov}
\end{figure*}

\XMM\ observed \psr\ on 2023 April 24--26 (ObsID 0900150101) for about
90~ks with EPIC in large window mode and using the thin optical filter.
Large window mode was chosen due to its higher time resolution, 47.7~ms
for pn, which allows for measurement of the 629~ms spin period of the pulsar.
We process EPIC MOS and pn observation data files (ODFs) using the
Science Analysis Software (SAS) 20.0.0 tasks \texttt{emproc} and
\texttt{epproc}, respectively.
To remove periods of background flaring, we extract single event
($\mbox{PATTERN}=0$), high energy ($>10$~keV for MOS and $10-12$~keV for pn)
light curves from which we determine low and steady count rate thresholds of
$0.2\mbox{ counts s$^{-1}$}$ for MOS and $0.4\mbox{ counts s$^{-1}$}$ for pn.
We use these thresholds to generate good time intervals (GTIs) with
\texttt{tabgtigen}, which we then use to produce flare-cleaned events
with \texttt{evtselect} and barycenter these events with \texttt{barycen}.
The resulting data have effective exposure times of 85~ks for MOS and
77~ks for pn.
We use both \texttt{edetect} and \texttt{wavdetect} to check the source
position and find consistent values within the uncertainty of the MOS
(and pn) on-axis point spread function of $4''$,
with R.A.$=08^{\rm h}37^{\rm m}57.7^{\rm s}$, decl.$=-24^\circ 54'30''$ (J2000).
This position is fully consistent with the radio position, and thus
we can positively associate our X-ray point source with that of \psr.
Figure~\ref{fig:fov} shows the MOS1, MOS2, and pn images,
and \psr\ is clearly evident in each dataset.

We consider the regions denoted in Figure~\ref{fig:fov} for extracting
MOS and pn source counts.
These are determined by first using \texttt{eregionanalyse}, which
indicates an optimum circular extraction radius of about $20''$.
Using \texttt{epatplot}, we do not find the source counts to be significantly
affected by pile-up.
We extract source and background counts for timing and spectral analysis
using $\mbox{PATTERN}\le12$ and $\le4$ for MOS and pn, respectively,
and $\mbox{FLAG}=0$ for spectra.
We apply light curve corrections using \texttt{epiclccorr}.
We calculate source and background areas and account for bad pixels and
chip gaps using \texttt{backscale}.
We then compute rmf and arf files.
We measure background-subtracted counts of 740, 720, and 4100
and count-rates of 0.0087, 0.0084, and 0.053~counts~s$^{-1}$
for MOS1, MOS2, and pn, respectively.
In order to improve statistics, we combine the MOS1 and MOS2 spectra using
\texttt{epicspeccombine}.
We bin spectra using \texttt{specgroup} to a minimum of 25 photons per bin
for the combined MOS spectrum and for the pn spectrum.

\section{Results} \label{sec:results}

\subsection{Radio spectral properties} \label{sec:radioresults}

\begin{figure}[hbt!]
\centering
\includegraphics[width=0.95\linewidth]{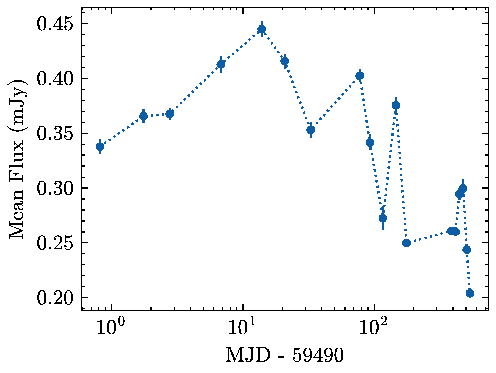}
\caption{Mean integrated flux density over time from Parkes Murriyang UWL
observations of \psr\ starting on 2021 October 3 (MJD~59490).
Errors shown are 1$\sigma$.}
\label{fig:flux_var}
\end{figure}

Figure~\ref{fig:flux_var} shows the mean integrated flux density of \psr\ over the Parkes observing campaign. This pulsar shows a significant amount of variation in the flux density over both short and long timescales, which is likely indicative of both refractive and diffractive scintillation from the intervening interstellar medium (ISM). On the other hand, neither the DM nor RM show any evidence for significant variations over the same timespan. Using a median $\rm RM = 360.72$~rad~m$^{-2}$ and a DM of 143.1~pc~cm$^{-3}$, we can derive an average magnetic field strength along the line of sight,
\begin{equation}
    \displaystyle B_{\parallel} = 1.232 \frac{\rm RM}{\rm DM} = 3.1 \, \mu {\rm G}.
\end{equation}

\begin{figure}[hbt!]
\centering
\includegraphics[width=0.95\linewidth]{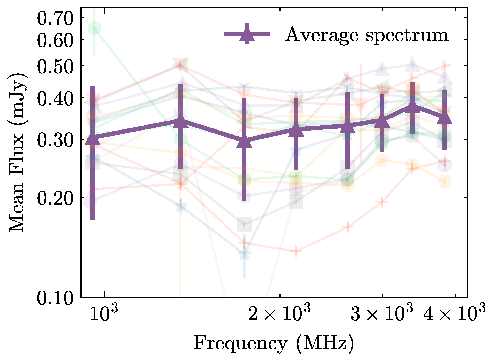}
\caption{Radio spectrum from Parkes Murriyang UWL observations of \psr.
Light colored points indicate mean integrated flux densities for each epoch of
observation, while dark purple points indicate the average at each frequency.
Errors shown are 1$\sigma$.}
\label{fig:spectrum}
\end{figure}

To measure the spectrum of \psr, we do not require the full native resolution of 1~MHz offered by the UWL receiver. Thus we downsampled our data to 8 frequency bins and calculated the mean integrated flux density in each bin. The spectrum calculated with this method for every epoch is shown in Figure~\ref{fig:spectrum} using transparent square markers. Given the large flux density variation as a function of time, we also averaged the per-epoch spectrum to get the mean spectrum, which is shown using solid purple markers in Figure~\ref{fig:spectrum}, where the uncertainty is given by the standard deviation of the measured flux density in each bin. A power-law fit ($S_{\nu} \propto \nu^{\alpha}$) to this mean spectrum yields a spectral index of $\alpha = 0.13 \pm 0.6$, though we note that this flat spectrum is likely due to effects of refractive scintillation, similar to that seen in, e.g., PSR~J1740+1000 \citep{j1740_mam}.

\subsection{Updated radio timing model} \label{sec:timing}

\begin{table}
\caption{Parameters for \psr}
\begin{tabular}{ll}
\toprule
\multicolumn{2}{c}{Dataset and model summary}\\
\midrule
Pulsar name                  \dotfill & \psr      \\
MJD range                    \dotfill & 59490---60025 \\
Data span (yr)               \dotfill & 1.46    \\
Number of TOAs               \dotfill & 18      \\
Solar system ephemeris       \dotfill & DE405      \\
PEPOCH \dotfill & 55588.628331 \\
\midrule
\multicolumn{2}{c}{Fit summary}\\
\midrule
Number of free parameters    \dotfill & 7      \\
RMS TOA residuals ($\mu s$) \dotfill & 56.07   \\
$\chi^2$                         \dotfill & 15.71    \\
Reduced $\chi^2$                 \dotfill & 1.57    \\
\midrule
\multicolumn{2}{c}{Measured Quantities} \\
\midrule
Right ascension (J2000) \dotfill &  08:37:57.748(2) \\
Declination (J2000) \dotfill &  $-$24:54:29.7(1) \\
F0 ($\mathrm{Hz}$)\dotfill &  $1.588788773(2)$ \\
F1 ($\mathrm{Hz\,s^{-1}}$)\dotfill &  $-8.80136(5) \times 10^{-13}$ \\
WXSIN\_0001 ($\mathrm{s}$)\dotfill &  $-0.0029(4)$ \\
WXCOS\_0001 ($\mathrm{s}$)\dotfill &  $-0.0015(2)$ \\
EFAC\dotfill &  $1.0(2)$ \\
\midrule
\multicolumn{2}{c}{Set Quantities} \\
\midrule
DM ($\mathrm{pc\,cm^{-3}}$)\dotfill &  143.100000 \\
WXEPOCH ($\mathrm{d}$)\dotfill &  59666.345636 \\
WXFREQ\_0001 ($\mathrm{d^{-1}}$)\dotfill &  0.001871 \\
\bottomrule
\end{tabular}
\label{table:timing_pars}
\end{table}

Using our radio data, we also generated a new timing solution for \psr{}. Since there was no significant variation in the DM relative to the fiducial DM used in the fold-mode observations, we held the DM fixed to this value. Given that there are no significant DM variations, we time and frequency scrunched each observation and, along with the standard profile from \citet{Poletal21}, derived a corresponding time-of-arrival (TOA) for each epoch using {\sc pat}.

\begin{figure}[hbt!]
\centering
\includegraphics[width=0.95\linewidth]{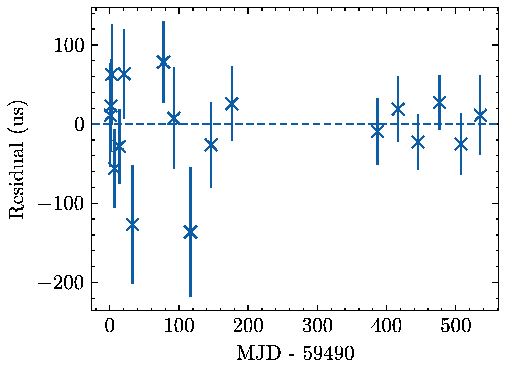}
\caption{Post-fit residuals, along with 1$\sigma$ uncertainties, from the Parkes timing campaign starting on 2021 October 3 (MJD~59490). Red noise is clearly visible, with the overall RMS of 100 $\mu$s in the residuals being comparable to the timing solution obtained in \citet{Poletal21}. We are unable to phase connect the TOAs from this observing campaign to those from \citet{Poletal21}.}
\label{fig:timing_res}
\end{figure}

The timing model and the corresponding residuals are shown in Table~\ref{table:timing_pars} and Figure~\ref{fig:timing_res}, respectively. We used PINT \citep{pint_software,Luoetal21} to time the pulsar, including modeling the white (EFAC only) and red noise \citep{pint_paper_2}. We used the timing model from \citet{Poletal21} to initialize our timing model after which the parameters were allowed to vary. We tried including more frequency derivatives, parallax, and proper motion to the timing model, but none of these parameters were found significant via an f-test.

Similar to \citet{Poletal21}, we notice significant red noise in the residuals for this pulsar. Without an explicit red noise process in the timing model, the red noise was absorbed in the fits to the sky-position of the pulsar, resulting in underestimating the uncertainty on the measured right ascension and declination for the pulsar by $\sim$4 and $\sim$6, respectively. To accurately reflect the uncertainties on these parameters, we include a red noise model with only a single harmonic, as determined using the WaveX model in PINT \citep{pint_paper_2}. We also attempted to phase connect the TOAs from these observations with those collected in \citet{Poletal21}, but we were unsuccessful due to the large red noise present in this pulsar.

\subsection{VLBA astrometric results} \label{sec:vlbiresults}

The 7--13$\sigma$ detections described in Section~\ref{sec:vlbiobs} enable us to measure the position of \psr\ with a nominal accuracy of 0.5--1~milliarcseconds in right ascension and 1--2~milliarcseconds in declination at each epoch. However, these statistical uncertainties do not encompass any residual calibration errors, which can be substantial, particularly at low observing elevations as is necessary for the VLBA at these declinations \citep[e.g.][]{Delleretal19}. In this instance, the significant angular broadening of the calibrator sources also means that longer integration times are required to achieve sufficient signal-to-noise ratio on the calibrators,  which further degrades the astrometric quality as the interpolation time increases.

\begin{figure}[hbt!]
\centering
\includegraphics[width=0.95\linewidth]{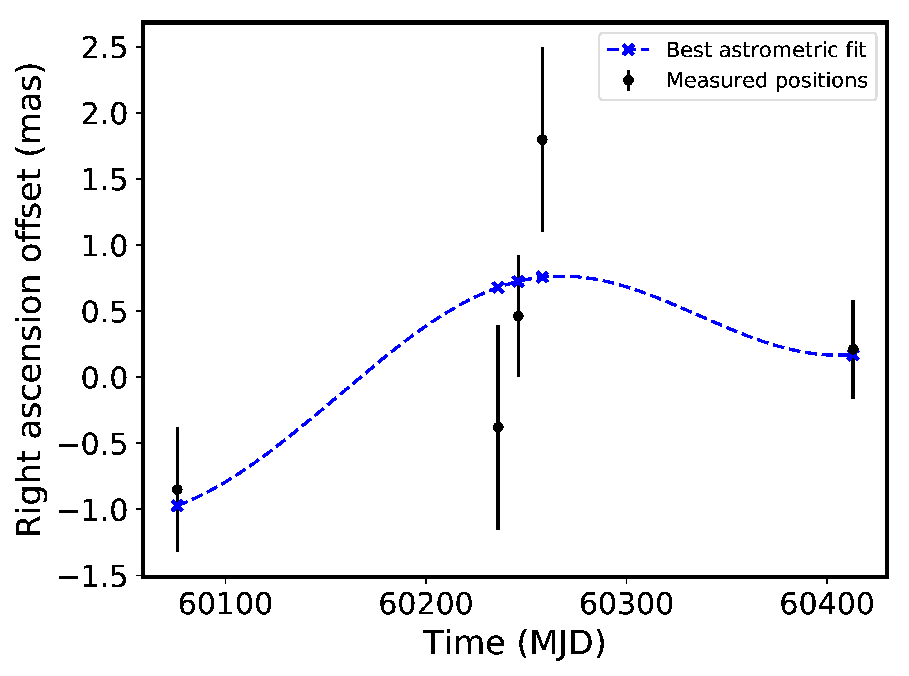} \\
\includegraphics[width=0.95\linewidth]{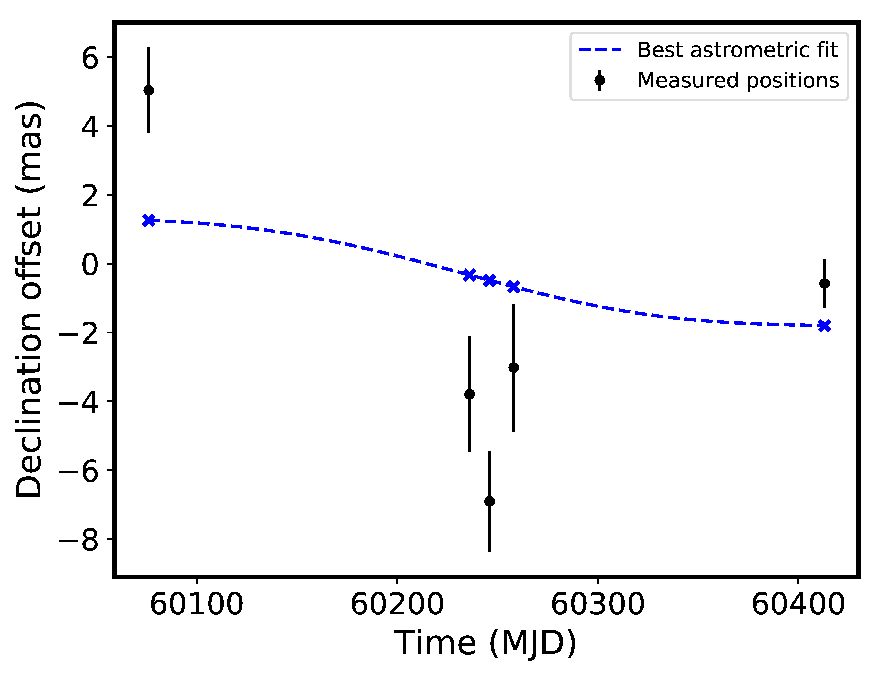}
\caption{The offset of \psr\ from the fitted reference position in right ascension (top) and declination (bottom) as a function of time. As discussed in the text, the error bars are from the image plane fit to the pulsar position only and are underestimated (particularly in declination).}
\label{fig:vlbi}
\end{figure}

In Table~\ref{tab:vlbifit}, we show the results of a straightforward least squares fit to the reference position, proper motion, and parallax for \psr\ based on the VLBA results, and we plot the model against the observed positions as a function of time in Figure~\ref{fig:vlbi}.However, significant unmodelled systematic errors are present in the data, as can be seen from the $\chi^2$ of the fit.  We discuss the degree to which these results (and a future extension to the VLBA astrometry) can constrain the distance to the pulsar in Section~\ref{sec:distance}. Finally, we note that the uncertainty in the reference position of the pulsar, which is not quoted in Table~\ref{tab:vlbifit}, is dominated by the uncertainty in the in-beam calibrator positions, which themselves were referenced to the primary phase calibrator located at a larger angular separation -- this uncertainty is likely to be on the order of 5 milliarcseconds in the current dataset.

\begin{table}
\caption{VLBA astrometric results for \psr}
\begin{tabular}{ll}
\toprule
\multicolumn{2}{c}{Fitted parameters}\\
\midrule
Right ascension (J2000)    \dotfill &  08:37:57.74880     \\
Declination (J2000) \dotfill & $-$24:54:29.941   \\
Proper motion, R.A. (mas yr$^{-1}$)    \dotfill &  $1.2\pm0.6$     \\
Proper motion, Decl. (mas yr$^{-1}$) \dotfill &   $-3.1\pm1.5$ \\
Parallax (mas) \dotfill & $0.59\pm0.24$\\
$\chi^2$                         \dotfill & 42.0    \\
Reduced $\chi^2$                 \dotfill & 8.4    \\
\midrule
\multicolumn{2}{c}{Set Quantities} \\
\midrule
Reference epoch (MJD) \dotfill &  60200 \\
\bottomrule
\end{tabular}
\label{tab:vlbifit}
\end{table}

\subsection{X-ray pulsation search and pulse profile} \label{sec:pulse}

\begin{figure}[hbt!]
\centering
\includegraphics[width=0.95\linewidth]{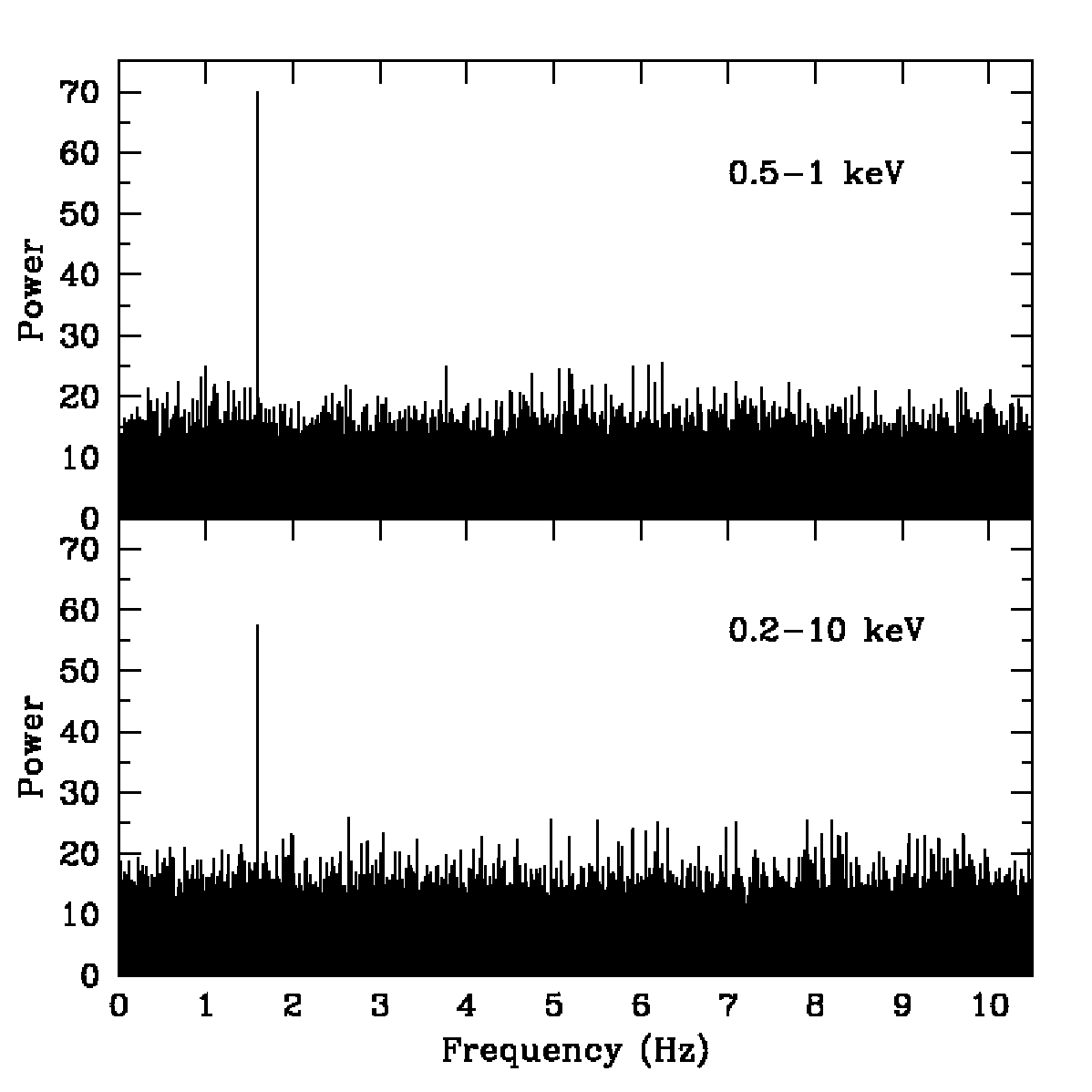}
\caption{Power spectra from pn data for 0.5--1~keV (top) and
0.2--10~keV (bottom).
The peak at 1.58845~Hz is the spin frequency of \psr.}
\label{fig:psearch}
\end{figure}

Figure~\ref{fig:psearch} shows power spectra of the pn data
generated using \texttt{powspec} in FTOOLS \citep{Blackburn95}.
A peak at 1.58845~Hz (629.545~ms) is evident and corresponds
to the spin frequency of \psr.
Note that \citet{Poletal21} measured a spin frequency of 1.58878890~Hz
($P=629.41024\mbox{ ms}$) and spin frequency time derivative of
$(-8.808\pm0.004)\times10^{-13}\mbox{ Hz s$^{-1}$}$
[$\dot{P}=(3.490\pm0.002)\times10^{-13}\mbox{ s s$^{-1}$}$]
at MJD~55588.628331,
such that the expected spin frequency at the time of the \XMM\
observation is $1.5884\pm0.0007$~Hz ($629.5\pm0.4$~ms),
which matches that from the power spectrum here.

\begin{figure}[hbt!]
\centering
\includegraphics[width=0.95\linewidth]{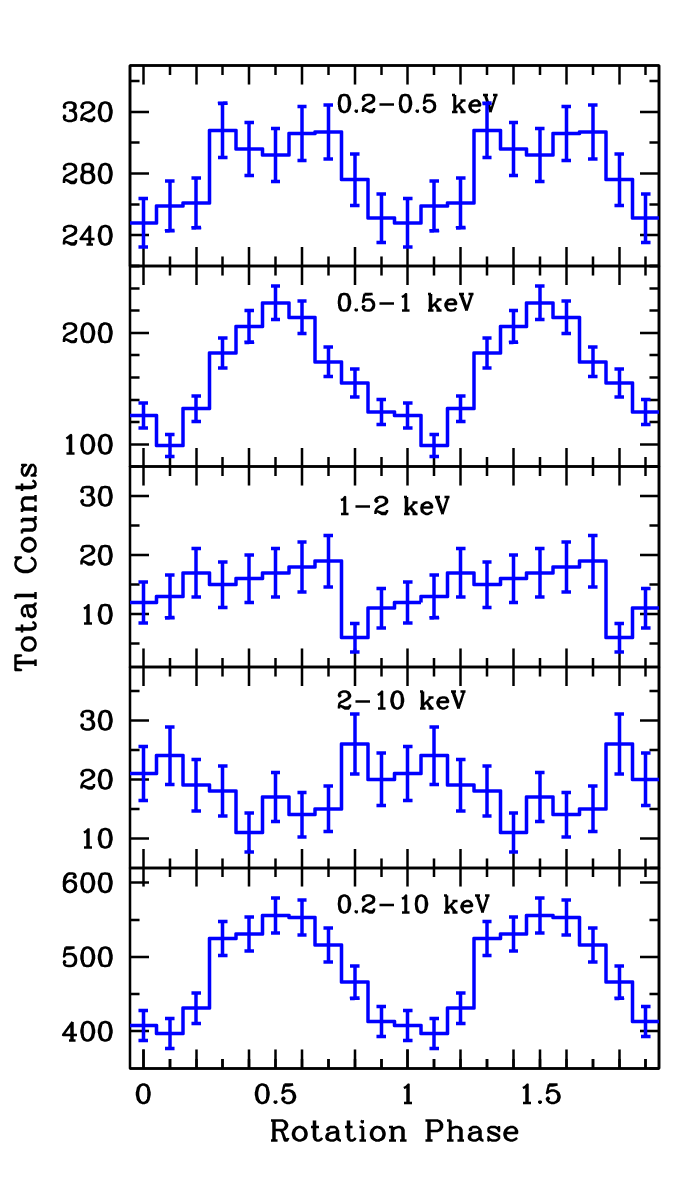}
\caption{Pulse profile of \psr, obtained by folding the pn data
in the energy ranges 0.2--0.5, 0.5--1, 1--2, 2--10, and 0.2--10~keV
at the 629.544~ms spin period.
Errors shown are 1$\sigma$.}
\label{fig:pp}
\end{figure}

Next, we perform acceleration searches using PRESTO \citep{Ransometal02}.
Data are folded at the candidate pulse frequency using \texttt{prepfold},
and a refined frequency is determined.
We find the strongest pulsations at a frequency of
1.58845213(9)~Hz or period of 629.54368(4)~ms (at MJD~60059.0334841)
in the energy range 0.5--1~keV.
We then use \texttt{photonphase} in PINT \citep{Luoetal21} to
calculate the rotation phase for each photon.
The resulting pulse profiles in various energy bands are shown in
Figure~\ref{fig:pp}.
Pulsations are only clearly evident at $E<1\mbox{ keV}$.
For the sinusoidal pulse profiles at 0.2--10~keV and 0.5--1~keV, 
the pulsed fractions [$=\mbox{(max-min)/(max+min)}$] are 17~percent
and 39~percent, respectively.

\subsection{X-ray spectral analysis} \label{sec:spectra}

\begin{figure}[hbt!]
\centering
\includegraphics[width=0.95\linewidth]{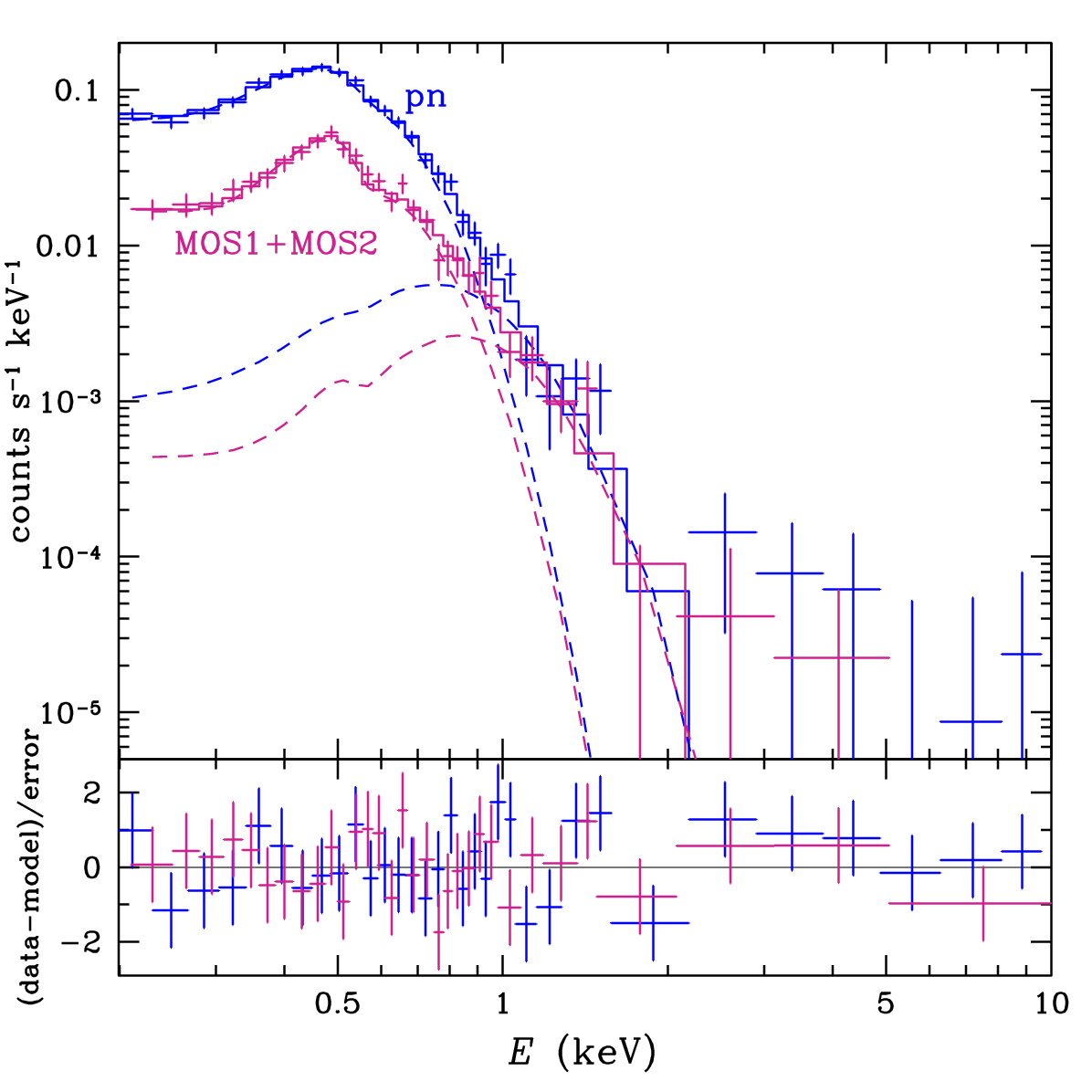}
\caption{Spectra of \psr\ from MOS1+MOS2 and pn data.
Top panel shows data with 1$\sigma$ errors (crosses) and
spectral model (dashed lines for each component).
Bottom panel shows $\chi^2=\mbox{(data-model)/error}$.
The spectral model shown here is a $B=10^{13}\mbox{ G}$
thin partially-ionized hydrogen atmosphere with two
temperature components;
a two temperature blackbody model yields similar results
(see text and Table~\ref{tab:spectra} for details).}
\label{fig:spectra}
\end{figure}

Figure~\ref{fig:spectra} shows the combined MOS (MOS1+MOS2) and pn spectra.
We perform model fits to these spectra using Xspec 12.14.0 \citep{Arnaud96}.
For the spectral model, we include several components.
First, we use \texttt{constant} to account for a possible instrumental
difference between MOS and pn spectral normalizations, and we fix its
value to 1 for the pn spectrum and allow it to vary for the combined
MOS spectrum.
Next, we include a component to account for photoelectric absorption by the
interstellar medium, i.e., \texttt{tbabs} with abundances from
\citet{Wilmsetal00} and cross-sections from \citet{Verneretal96}.
To model the intrinsic spectrum of \psr, we first consider a single
component comprising a power law (PL; \texttt{powerlaw}),
blackbody (BB; \texttt{bbodyrad}), or neutron star atmosphere.
For the last, we use a
magnetic partially ionized hydrogen atmosphere model (\texttt{nsmaxg};
\citealt{Hoetal08,Ho14}) and fix the model parameters of neutron star
mass and radius to $M=1.4\,\Msun$ and $R=10\mbox{ km}$, respectively.
We also tried
a non-magnetic fully ionized hydrogen atmosphere model (\texttt{nsatmos};
\citealt{Heinkeetal06}),
a non-magnetic partially ionized hydrogen, helium, or carbon atmosphere model
(\texttt{nsx}; \citealt{HoHeinke09}),
and a magnetic partially ionized carbon, oxygen, or neon atmosphere model
\citep{MoriHo07},
but none of these provide good fits and thus will not be discussed further.

The results of fits with a single component model for the intrinsic
spectrum of \psr, i.e., either power law, blackbody, or atmosphere model,
are given in Table~\ref{tab:spectra}.
We see that the power law model not only has the worst fit of the three,
but the best-fit power law index $\Gamma\approx 8$ is unrealistic.
Meanwhile, thermal models provide better fits if the distance to the
pulsar is taken to be less than a few kpc.
For example, in order for the best-fit blackbody emission radius $\Rinfty$ [$=R(1-2GM/c^2R)^{-1/2}$] to
be less than $\sim19\mbox{ km}$ (corresponding to a maximum neutron star radius $R\approx 13-15\mbox{ km}$), then the
distance must be $d<4\mbox{ kpc}$ (after accounting for gravitational
redshift).
Similarly assuming that the entire stellar surface emits at a single
uniform temperature, the best-fit atmosphere model gives a distance
of only 90~pc.
We note that \texttt{nsmaxg} models with $B=2$, 4, or $7\times10^{12}\mbox{ G}$
yield fits that are about as good as that of $10^{13}\mbox{ G}$ and
have best-fit parameters $\NH$ increasing, $\Teff$ decreasing, and
$d$ decreasing with lower values of $B$,
while fits are poor for models with lower and higher magnetic fields,
i.e., $10^{10}-10^{12}\mbox{ G}$ and $(2-3)\times10^{13}\mbox{ G}$.
The absorbed 0.3--10~keV flux is $7.2\times10^{-14}\mbox{ erg s$^{-1}$ cm$^{-2}$}$, so that the unabsorbed 0.3--10~keV X-ray luminosity $\Linfty_{\rm X}\sim3.3\times10^{31}\mbox{ erg s$^{-1}$}(d/0.9\mbox{ kpc})^2$
and bolometric luminosity $\Linfty\sim7.6\times10^{31}\mbox{ erg s$^{-1}$}(d/0.9\mbox{ kpc})^2$.

\begin{table*}[hbt!]
\caption{Results of spectral modeling}
\label{tab:spectra}
\begin{tabular}{llllllll}
\toprule
\headrow & PL$^{\rm a}$ & BB$^{\rm a}$ & NSMAXG$^{\rm b}$ & BB+PL$^{\rm a}$ & NSMAXG+PL$^{\rm a,b}$ & BB+BB$^{\rm a}$ & NSMAXG+NSMAXG$^{\rm b}$ \\
\midrule
MOS/pn normalization & $1.10\pm0.04$ & $1.14\pm0.04$ & $1.14\pm0.04$ & $1.13\pm0.04$ & $1.13\pm0.04$ & $1.13\pm0.04$ & $1.15\pm0.04$ \\
$\NH$ ($10^{21}\mbox{ cm$^{-2}$}$) & $3.7\pm0.2$ & $1.6\pm0.1$ & $2.6\pm0.2$ & $2.1^{+0.3}_{-0.2}$ & $2.7\pm0.2$ & $2.0\pm0.2$ & $1.8\pm0.2$ \\
$k\Tinfty$ or $k\Teff$ (eV) & & $78\pm2$ & $27\pm1$ & $69\pm3$ & $27\pm1$ & $70\pm3$ & $50^{+4}_{-3}$ \\
$\Rinfty/d$ & & $4.4^{+0.6}_{-0.5}\mbox{ km}/1\mbox{ kpc}$ & $13\mbox{ km}/90\pm20\mbox{ pc}$ & $7.4^{+1.7}_{-1.3}\mbox{ km}/1\mbox{ kpc}$ & $13\mbox{ km}/80\pm20\mbox{ pc}$ & $7.5^{+1.9}_{-1.3}\mbox{ km}/1\mbox{ kpc}$ & $13\mbox{ km}/1.1^{+0.4}_{-0.3}\mbox{ kpc}$ \\
$\Gamma$ & $8.2\pm0.2$ & & & $5.7^{+0.8}_{-0.9}$ & $1.9^{+3.4}_{-1.7}$ & & \\
PL normalization ($10^{-6}$) & $13.4\pm0.6$ & & & $5\pm1$ & $0.34^{+1.16}_{-0.31}$ & & \\
$k\Tinftyb$ (eV) & & & & & & $160^{+30}_{-20}$ & $180^{+60}_{-30}$ \\
$\Rinftyb$ (m) & & & & & & $150^{+140}_{-70}$ & $120^{+50}_{-20}$ \\
\midrule
$\chi^2/\mbox{dof}$ & 83.8/61 & 72.8/61 & 61.4/61 & 49.8/59 & 59.5/59 & 47.4/59 & 45.8/59 \\
\bottomrule
\end{tabular}
\begin{tablenotes}[hang]
\item[]Errors are $1\sigma$.
\item[a]Xspec models: PL=\texttt{powerlaw}, BB=\texttt{bbodyrad}.
\item[b]Neutron star mass $M=1.4\,\Msun$ and radius $R=10\mbox{ km}$ are fixed, such that a redshift of 1.3 is used to convert between unredshifted and redshifted values.
Distance is allowed to vary with fit, while fraction of surface emitting is fixed to 1.
Magnetic field $B=10^{13}\mbox{ G}$ is assumed (see text for other values).
\end{tablenotes}
\end{table*}

Next, we consider two component models for fitting the measured spectra,
on top of \texttt{constant} and \texttt{tbabs},
in particular, two power laws (PL+PL), a blackbody and a power law
(BB+PL), or two blackbodies (BB+BB).
For two power laws, the best-fit does not yield an improvement in $\chi^2$,
and the second power law only has an upper limit on its normalization.
Meanwhile, the best-fit results for BB+PL and BB+BB are given in
Table~\ref{tab:spectra}.
While we see an improvement in $\chi^2$ for BB+PL compared to only
a power law or only a blackbody, with the power law providing the
primary flux at $>1\mbox{ keV}$, the power law index is still a  high
$\Gamma>5$.
Meanwhile, two blackbodies yield a much improved fit ($\Delta\chi^2=25.4$
for two extra degrees of freedom, so that a f-test gives a probability
of $3\times10^{-6}$).  The second blackbody with a higher temperature and
small emitting region could be interpreted as emission from a hot spot on
the neutron star surface that rotates to
generate the pulsations detected and described in Section~\ref{sec:pulse}.
The size of the colder blackbody limits the distance to \psr\ to be
$d<1.7\mbox{ kpc}$.
We note for completeness that we also fit the spectra with \texttt{nsmaxg}+PL,
with best-fit results shown in Table~\ref{tab:spectra}.
The small improvement of the best-fit over only \texttt{nsmaxg}
shows that the addition of the power law (to better fit the spectra
at high energies) is not really needed.
This difference with the BB+PL results is because neutron star
atmosphere model spectra are harder at high energies than blackbody
spectra at the same temperature due to the atmospheric opacity decreasing with
energy, such that higher energy photons come from deeper, hotter
layers of the atmosphere (see, e.g., \citealt{Pavlovetal95}).
This also implies that fitting spectra with a single blackbody gives
a higher temperature and smaller emitting region compared to fitting
with an atmosphere model, as evident from the results given in Table~\ref{tab:spectra}.

Finally, we fit the X-ray spectra using a two component neutron star
atmosphere model, like the two component blackbody model described
above.  Such a model with emission from a small hot magnetic polar cap
and emission from the remaining cooler surface can produce rotation
modulated pulsations.  A more robust distance estimate can also be determined.
Since a single atmosphere model provides a good fit to the observed
spectra without a need for a second component and guided by the good
fit of the two blackbody model, the model we use here is a thin
atmosphere model.
A thin atmosphere is one whose physical thickness is such that the
atmosphere is optically thick ($\tau>1$) at low energies and
optically thin ($\tau<1$) at high energies.
Thus the observed high energy photons arise from a condensed surface
beneath the gaseous atmosphere, and this condensed surface emits
blackbody-like emission (away from features such as electron and ion
cyclotron resonances;
see, e.g., \citealt{vanAdelsbergetal05,Potekhinetal12,Potekhin14}).
Such thin atmospheres were first used to model the spectrum of the
strongly magnetic ($10^{12}-10^{13}\mbox{ G}$) neutron stars
RX~J0720.4$-$3125 \citep{Motchetal03} and
RX~J1856.5$-$3754 \citep{Ho07,Hoetal07}.
Here we use \texttt{nsmaxg} models with condensed surface
spectra from \citet{vanAdelsbergetal05}.
The best-fit results are shown in Table~\ref{tab:spectra} and
Figure~\ref{fig:spectra}
(for a Thomson depth of 0.7, corresponding to an atmosphere thickness of 3.8~cm).
We see that the best-fit is as good as the two blackbody fit and
a better fit than that of a single atmosphere.
Taking the cool component as being due to the entire neutron
star surface, we estimate the distance to \psr\ to be
$d\sim1\mbox{ kpc}$.
Results for somewhat larger depths than 0.7 are roughly
within the uncertainties given in Table 3 and tend toward lower
temperatures and a smaller distance, while smaller depths produce
essentially blackbodies in X-rays.

We note here that the best-fit interstellar absorption $\NH$ is in
the range $\sim(1.8-2)\times10^{21}\mbox{ cm$^{-2}$}$.
These are a factor of 2--3 larger than the Galactic column density of
$\NH=7.3\times10^{20}\mbox{ cm$^{-2}$}$ in the direction of \psr\
\citep{Hi4pi16}.
On the other hand, the empirical relation between $\NH$ and dispersion
measure from \citet{Heetal13} yields
$\NH=4\times10^{21}\mbox{ cm$^{-2}$}$ for the pulsar's dispersion measure of
$143.1\mbox{ pc cm$^{-3}$}$.
The relatively large $\NH$ from our spectral fits, as well as the high DM,
could be indicating significant material in our line-of-sight from,
for example, a supernova remnant.
This material may also contribute to the strong scattering seen in radio
observations.

\section{Discussion} \label{sec:discuss}

\subsection{Implications on distance to and kinematics \label{sec:distance}}

As shown in Section~\ref{sec:timing}, while we are able to produce an updated timing solution, we are unable to phase connect the new observations with those taken in \citet{Poletal21}. Consequently, we are unable to measure a timing parallax and infer the distance to the pulsar.
While we detect the pulsar in a number of the VLBA observations, the signal-to-noise ratio and hence astrometric precision is lower than expected. As shown in Table~\ref{tab:vlbifit}, the five available position measurements imply a best-fit parallax of 0.6~milliarcseconds and hence best-fit distance of 1.7~kpc. The $\chi^2$ value of 42 for 5 degrees of freedom shows that the parallax uncertainty of 0.24~milliarcseconds is likely to be significantly underestimated, perhaps by a factor of three or more. Accordingly, we cannot measure the distance using the current dataset but can only say that very nearby distances are increasingly disfavoured. For instance, even assuming that the uncertainty is underestimated by a factor of 5, a distance below 330~pc would be excluded at 95\% confidence.  An extended VLBA astrometric campaign would rapidly improve the distance constraints in three ways: 1) the additional data points would improve the statistical precision of the astrometric fit, 2) the larger ensemble of data points would allow the true distribution of the systematic error terms to be better estimated, and 3) the improved time baseline would greatly reduce the covariance between proper motion (in right ascension) and parallax, reducing the parallax uncertainty.

The current, uncertain lower limit to the distance based on the VLBA astrometry can still be compared to constraints on the pulsar distance based on fits to its X-ray spectrum. As described in Section~\ref{sec:spectra}, all the spectral models that provide a good fit to the data imply a distance less than that predicted from the NE2001 and YMW16 dispersion measure models. The X-ray spectra give a distance to \psr{} of $d \lesssim 1$~kpc, similar to the indirectly estimated distances found in \citet{Poletal21}. This is not in tension with the VLBA results, although these favor a distance towards the upper end of this range.

Finally, we note that the VLBA proper motion implies a relatively low transverse velocity for the system. While the exact value depends on the assumed distance, a reasonable maximum can be estimated by taking a distance at the upper end of the range allowed by the X-ray spectral modelling and assuming the same factor-of-five increase in the nominal VLBA uncertainties. In this case, the proper motion upper limit is $\lesssim$10 mas yr$^{-1}$, corresponding to a transverse velocity of $\lesssim$50 km s$^{-1}$, considerably slower than most pulsars \citep[e.g.][]{Igoshev2020}.

\subsection{Neutron star cooling implications} \label{sec:nscool}

\begin{figure*}[hbt!]
\centering
\includegraphics[width=0.45\linewidth]{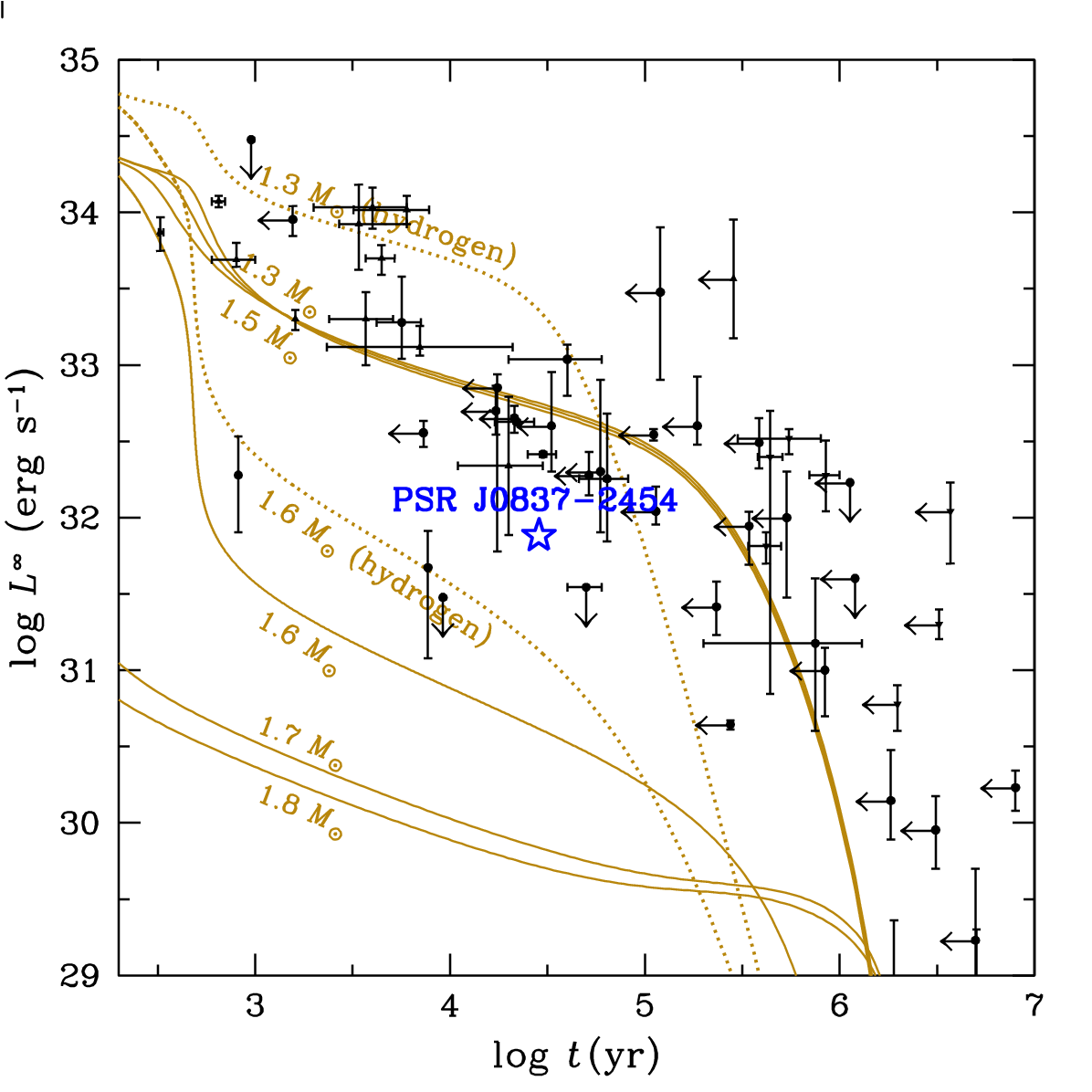}
\includegraphics[width=0.45\linewidth]{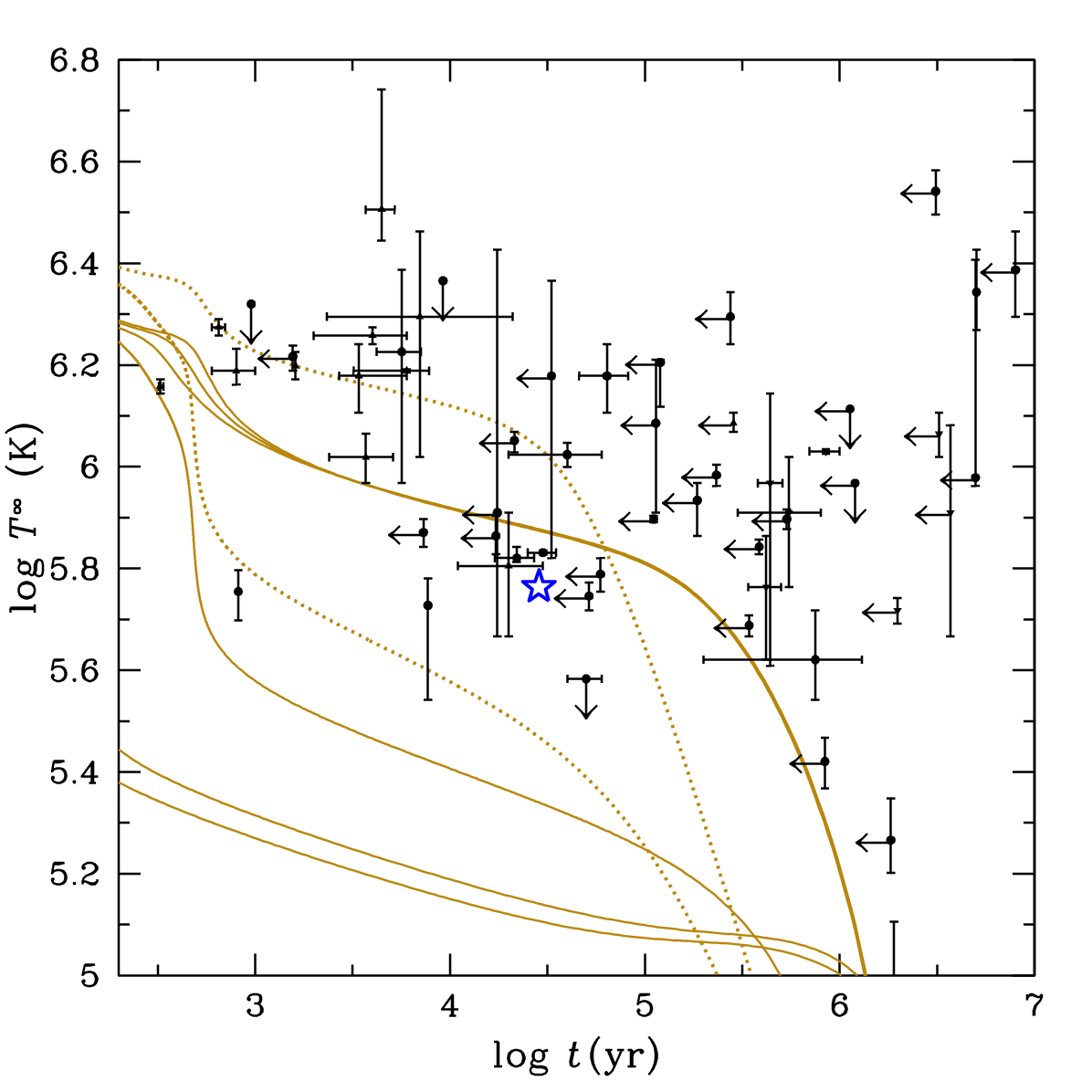}
\caption{Observed luminosity $L^\infty$ (left) and temperature $T^\infty$
(right) as functions of neutron star age.
Data points are from \citet{Potekhinetal20}
(see also https://www.ioffe.ru/astro/NSG/thermal/),
except the stars which denote \psr.
Solid lines show cooling curves from neutron star cooling simulations
for neutron star masses $M=1.3$, 1.4, 1.5, 1.6, 1.7, and $1.8\,\Msun$
(from top to bottom)
using the BSk24 nuclear equation of state and an iron envelope and
including neutron superfluidity in the crust and core and proton
superconductivity in the core, while dashed lines are for a hydrogen envelope;
note that direct Urca cooling becomes active for BSk24 at $M>1.59\,\Msun$
(see text for details).}
\label{fig:nscool}
\end{figure*}

For the purposes of the discussion in this section, we assume a distance
to \psr\ of 0.9~kpc, derived from associating diffuse emission detected
by GLEAM to be the supernova remnant left over from the formation of the
pulsar.
At this distance, the measured X-ray flux results in a thermal luminosity
of $\Linfty\sim7.6\times10^{31}\mbox{ erg s$^{-1}$ }(d/0.9\mbox{ kpc})^2$.
The best-fit spectral model of a neutron star $B=10^{13}\mbox{ G}$
atmosphere yields a surface temperature of $\Teff=5.8\times10^5\mbox{ K}$.
Both the luminosity and temperature are low compared to most neutron stars
of age $<10^5\mbox{ yr}$, which have
$\Linfty>10^{32}\mbox{ erg s$^{-1}$}$ and $\Teff>6\times10^5\mbox{ K}$,
as shown in Figure~\ref{fig:nscool}.
On the other hand, the $\Linfty$ and $\Teff$ of \psr\ are comparable to
those of three pulsars of approximately similar age, i.e.,
PSR~J0007+7303 ($\Linfty<3\times10^{31}\mbox{erg s$^{-1}$}$, $\log\Teff<6.4$,
and age$\sim9\mbox{ kyr}$; \citealt{Caraveoetal10,Martinetal16,Potekhinetal20}),
PSR~B1727$-$47 ($\Linfty<3\times10^{31}\mbox{erg s$^{-1}$}$, $\log\Teff<5.6$,
and age$\sim50\mbox{ kyr}$; \citealt{Shterninetal19,Potekhinetal20}), and
PSR~B2334+61 ($\Linfty\sim5\times10^{31}\mbox{ erg s$^{-1}$}$,
$\log\Teff\approx5.6$, and age$\sim7.7\mbox{ kyr}$;
\citealt{Yaruyanikeretal04,Mcgowanetal06,Potekhinetal20}).
It is worth noting that the ages of these three pulsars are estimated
from their associated supernova remnant and each age is lower than the
characteristic spin-down age of the pulsar
(i.e., $\tau_{\rm c}=14$~kyr for PSR~J0007+7303, 80~kyr for PSR~B1727$-$47, and
41~kyr for PSR~B2334+61).
This may indicate that the age of \psr\ is lower than its characteristic
age of 28.6~kyr.

The luminosity and temperature of most neutron stars can be explained by
``slow'' neutrino cooling predominately from the modified Urca processes
and enhanced by Cooper pairing processes, with variations due to each
neutron star's envelope composition and age (\citealt{Gusakovetal04,Pageetal04};
see, e.g., \citealt{Potekhinetal15}, for review).
An envelope composed of light elements is more effective at conducting
interior heat and makes the surface temperature closer to that of the
interior temperature and thus higher than the surface temperature for an
envelope composed of heavy elements.
The age determines how much of the crust and core neutrons and core protons
are in a superfluid and superconducting state, respectively, in particular
the ratio between the interior temperature $T$ relative to the
(theoretically-uncertain) critical temperatures $\Tc$ for neutron
superfluidity and proton superconductivity.
When $T/\Tc<1$, cooling processes that depend on matter being in a
non-superfluid state are suppressed, while cooling by Cooper pair breaking
and formation of superfluid matter becomes effective.

The low luminosity and temperature of neutron stars like \psr\
point to neutrino cooling processes in these stars that operate much faster
than modified Urca and Cooper pairing.
Neutrino emission by direct Urca reactions is such a fast cooling process,
but it requires a high proton fraction ($\gtrsim 0.11$) in the neutron star
core \citep{Lattimeretal91,PageApplegate92}.
Such high proton fractions are only reached in neutron stars of high mass,
and some theoretical nuclear equations of state (EOSs) do not even yield
the necessary proton fractions at their predicted maximum mass.
For example, direct Urca cooling can occur in neutron stars with mass
$M>1.96\,\Msun$ for the APR EOS \citep{Akmaletal98} and $M>1.59\,\Msun$
for the BSk24 EOS but does not occur for the BSk26 EOS \citep{Pearsonetal18}
and the SLy4 EOS \citep{DouchinHaensel01}.

To illustrate the above, we perform neutron star cooling simulations
using the neutron star cooling code from \citet{Hoetal12,Hoetal12b},
with revised neutrino emissivities for modified Urca \citep{Shterninetal18},
plasmon decay \citep{KantorGusakov07}, and
neutron triplet pairing \citep{Leinson10} and
superfluid suppression factors \citep{Gusakov02}
(see \citealt{PotekhinChabrier18}, for discussion of some of the above).
We use the BSk24 EOS \citep{Pearsonetal18} and parameterization of
superfluid and superconductor energy gaps from \citet{Hoetal15} for the
SFB neutron singlet model \citep{Schwenketal03},
TToa neutron triplet model \citep{TakatsukaTamagaki04}, and
CCDK proton singlet model \citep{Chenetal93}.
It is important to note that the resulting cooling evolutions vary greatly
on the assumed EOS and superfluid and superconductor energy gaps;
nevertheless our conclusions on direct Urca cooling and neutron star mass
are still valid across most of these theoretical model variations.

The luminosity and temperature evolutions from our cooling simulations are
shown in Figure~\ref{fig:nscool} for an iron envelope and neutron star masses
$M=1.3,1.4,\ldots,1.8\,\Msun$,
as well as for a hydrogen envelope and $M=1.3$ and $1.6\,\Msun$.
We see that cooling curves for masses below the direct Urca threshold of
$1.59\,\Msun$ (and for same envelope composition) do not differ much from
each other except at early times.
These slow cooling curves can explain the luminosity and temperatures (or
their limits) of most observed neutron stars given their ages (or limits on
their ages) when we allow some to have a light element envelope.
Meanwhile, some of the more luminous and hotter sources are magnetars, which
are heated internally by magnetic field decay that is not included in our
present calculations but are included in works such as
\citet{Kaminkeretal06,Hoetal12,Viganoetal13,SkiathasGourgouliatos24}.
On the other hand, we see that cooling curves for masses above the direct
Urca threshold decrease to much lower luminosities and temperatures earlier
and that fast cooling curves such as these can match the measurements of
neutron stars such as \psr.
Therefore this indicates that, not only does \psr\ have a high mass, but also
the true nuclear equation of state should be one that can produce stable neutron
stars with a high enough proton fraction in the core to allow direct Urca
neutrino emission processes.

\section{Summary} \label{sec:summary}

\psr\ is a young radio pulsar at a high Galactic latitude with an
inferred magnetic field of $\sim 10^{13}\mbox{ G}$ but uncertain distance.
Using \XMM, we detected its spin pulsations and thus identify it as a
X-ray pulsar for the first time.
The X-ray pulse profile of \psr\ is broad, sinusoidal, strongest at
low energies, and probably produced by emission from a rotating hot
spot on a cooler surface based on properties of the spectrum.
The X-ray spectrum can be described as a soft thermal spectrum
that can be well-fit by a predominately cool component with a blackbody
temperature of 70~eV or atmosphere temperature of 50~eV, as well as a
160--180~eV hot spot component.
Results of the spectral analysis also indicate \psr\ is at a distance
of $\lesssim 1\mbox{ kpc}$, which is much less than that inferred from
the dispersion measure of the radio data but in agreement with other
information from radio and optical.  VLBA astrometry presently only provide a lower limit to the distance, favouring a distance $\gtrsim330$~pc.

Assuming a distance of 0.9~kpc, the measured luminosity of
$7.6\times10^{31}\mbox{ erg s$^{-1}$}(d/\mbox{0.9 kpc})^2$ implies \psr\
is one of the coldest young neutron stars known.
Neutron star cooling theory indicates that such a low luminosity and temperature
are only achievable if \psr\ has a mass high enough to activate fast direct
Urca neutrino emission processes.  Such a condition requires a nuclear
EOS that can reach a high proton fraction in the stellar core of stable
high mass neutron stars.

The distance to \psr{} remains the primary uncertainty regarding this
source. We are unable to measure a timing parallax through our Parkes data due to the high red noise present in this pulsar's emission, while the angular broadening of the pulsar in VLBA imaging led to lower astrometric precision than originally anticipated.
An extension of the VLBA observations presented here could improve the precision of the VLBA parallax by a factor of several and provide a distance measurement, as opposed to the current lower limit.
Confirmation of a supernova remnant association would also help with the distance and provide an estimate of the pulsar's true age.

\begin{acknowledgement}
The authors are grateful to Harsha Blumer, Natasha Hurley-Walker, and Simon Johnston for discussions and support.
N.P. thanks Rahul Sengar, David Kaplan, and Michael Lam for useful suggestions
and discussion regarding processing the radio timing data.
The authors thank the anonymous referee for comments that led to improvements
in the paper.
W.C.G.H. appreciates the use of computer facilities at the Kavli Institute
for Particle Astrophysics and Cosmology.
\end{acknowledgement}

\paragraph{Funding Statement}
W.C.G.H. acknowledges support through grant 80NSSC24K0329 from NASA.
N.P was supported through NSF Physics Frontiers Center award PFC-2020265.

\paragraph{Competing Interests}
None

\paragraph{Data Availability Statement}
The data underlying this article will be shared on reasonable request
to the corresponding author.

\bibliography{psrj0837}

\begin{thebibliography}{}
\expandafter\ifx\csname natexlab\endcsname\relax\def\natexlab#1{#1}\fi

\bibitem[{{Abhishek} {et~al.}(2022){Abhishek}, {Tanushree}, {Hegde}, \&
  {Konar}}]{Abhisheketal22}
{Abhishek}, Malusare, N., {Tanushree}, N., {Hegde}, G., \& {Konar}, S. 2022,
  Journal of Astrophysics and Astronomy, 43, 75

\bibitem[{{Akmal} {et~al.}(1998){Akmal}, {Pandharipande}, \&
  {Ravenhall}}]{Akmaletal98}
{Akmal}, A., {Pandharipande}, V.~R., \& {Ravenhall}, D.~G. 1998, \prc, 58, 1804

\bibitem[{{Arnaud}(1996)}]{Arnaud96}
{Arnaud}, K.~A. 1996, in Astronomical Society of the Pacific Conference Series,
  Vol. 101, Astronomical Data Analysis Software and Systems V, ed. G.~H.
  {Jacoby} \& J.~{Barnes}, 17

\bibitem[{{Becker}(2009)}]{Becker09}
{Becker}, W. 2009, in Astrophysics and Space Science Library, Vol. 357,
  Astrophysics and Space Science Library, ed. W.~{Becker}, 91

\bibitem[{{Blackburn}(1995)}]{Blackburn95}
{Blackburn}, J.~K. 1995, in Astronomical Society of the Pacific Conference
  Series, Vol.~77, Astronomical Data Analysis Software and Systems IV, ed.
  R.~A. {Shaw}, H.~E. {Payne}, \& J.~J.~E. {Hayes}, 367

\bibitem[{{Burke-Spolaor} {et~al.}(2011){Burke-Spolaor}, {Bailes}, {Johnston},
  {Bates}, {Bhat}, {Burgay}, {D'Amico}, {Jameson}, {Keith}, {Kramer}, {Levin},
  {Milia}, {Possenti}, {Stappers}, \& {van Straten}}]{Burkespolaoretal11}
{Burke-Spolaor}, S., {Bailes}, M., {Johnston}, S., {et~al.} 2011, \mnras, 416,
  2465

\bibitem[{{Caraveo} {et~al.}(2010){Caraveo}, {De Luca}, {Marelli}, {Bignami},
  {Ray}, {Saz Parkinson}, \& {Kanbach}}]{Caraveoetal10}
{Caraveo}, P.~A., {De Luca}, A., {Marelli}, M., {et~al.} 2010, \apjl, 725, L6

\bibitem[{{Chen} {et~al.}(1993){Chen}, {Clark}, {Dav{\'e}}, \&
  {Khodel}}]{Chenetal93}
{Chen}, J.~M.~C., {Clark}, J.~W., {Dav{\'e}}, R.~D., \& {Khodel}, V.~V. 1993,
  \nphysa, 555, 59

\bibitem[{{Cordes} \& {Lazio}(2002)}]{CordesLazio02}
{Cordes}, J.~M., \& {Lazio}, T.~J.~W. 2002, arXiv e-prints, astro

\bibitem[{{Deller} {et~al.}(2011){Deller}, {Brisken}, {Phillips}, {Morgan},
  {Alef}, {Cappallo}, {Middelberg}, {Romney}, {Rottmann}, {Tingay}, \&
  {Wayth}}]{Delleretal11}
{Deller}, A.~T., {Brisken}, W.~F., {Phillips}, C.~J., {et~al.} 2011, \pasp,
  123, 275

\bibitem[{{Deller} {et~al.}(2019){Deller}, {Goss}, {Brisken}, {Chatterjee},
  {Cordes}, {Janssen}, {Kovalev}, {Lazio}, {Petrov}, {Stappers}, \&
  {Lyne}}]{Delleretal19}
{Deller}, A.~T., {Goss}, W.~M., {Brisken}, W.~F., {et~al.} 2019, \apj, 875, 100

\bibitem[{{Ding} {et~al.}(2023){Ding}, {Deller}, {Stappers}, {Lazio}, {Kaplan},
  {Chatterjee}, {Brisken}, {Cordes}, {Freire}, {Fonseca}, {Stairs},
  {Guillemot}, {Lyne}, {Cognard}, {Reardon}, \& {Theureau}}]{Dingetal23}
{Ding}, H., {Deller}, A.~T., {Stappers}, B.~W., {et~al.} 2023, \mnras, 519,
  4982

\bibitem[{{Douchin} \& {Haensel}(2001)}]{DouchinHaensel01}
{Douchin}, F., \& {Haensel}, P. 2001, \aap, 380, 151

\bibitem[{{Gaustad} {et~al.}(2001){Gaustad}, {McCullough}, {Rosing}, \& {Van
  Buren}}]{Gaustadetal01}
{Gaustad}, J.~E., {McCullough}, P.~R., {Rosing}, W., \& {Van Buren}, D. 2001,
  \pasp, 113, 1326

\bibitem[{{Gusakov}(2002)}]{Gusakov02}
{Gusakov}, M.~E. 2002, \aap, 389, 702

\bibitem[{{Gusakov} {et~al.}(2004){Gusakov}, {Kaminker}, {Yakovlev}, \&
  {Gnedin}}]{Gusakovetal04}
{Gusakov}, M.~E., {Kaminker}, A.~D., {Yakovlev}, D.~G., \& {Gnedin}, O.~Y.
  2004, \aap, 423, 1063

\bibitem[{{He} {et~al.}(2013){He}, {Ng}, \& {Kaspi}}]{Heetal13}
{He}, C., {Ng}, C.~Y., \& {Kaspi}, V.~M. 2013, \apj, 768, 64

\bibitem[{{Heinke} {et~al.}(2006){Heinke}, {Rybicki}, {Narayan}, \&
  {Grindlay}}]{Heinkeetal06}
{Heinke}, C.~O., {Rybicki}, G.~B., {Narayan}, R., \& {Grindlay}, J.~E. 2006,
  \apj, 644, 1090

\bibitem[{{Hern{\'a}ndez-Pajares} {et~al.}(2009){Hern{\'a}ndez-Pajares},
  {Juan}, {Sanz}, {Orus}, {Garcia-Rigo}, {Feltens}, {Komjathy}, {Schaer}, \&
  {Krankowski}}]{igsg_1}
{Hern{\'a}ndez-Pajares}, M., {Juan}, J.~M., {Sanz}, J., {et~al.} 2009, Journal
  of Geodesy, 83, 263

\bibitem[{{HI4PI Collaboration} {et~al.}(2016){HI4PI Collaboration}, {Ben
  Bekhti}, {Fl{\"o}er}, {Keller}, {Kerp}, {Lenz}, {Winkel}, {Bailin},
  {Calabretta}, {Dedes}, {Ford}, {Gibson}, {Haud}, {Janowiecki}, {Kalberla},
  {Lockman}, {McClure-Griffiths}, {Murphy}, {Nakanishi}, {Pisano}, \&
  {Staveley-Smith}}]{Hi4pi16}
{HI4PI Collaboration}, {Ben Bekhti}, N., {Fl{\"o}er}, L., {et~al.} 2016, \aap,
  594, A116

\bibitem[{{Ho}(2007)}]{Ho07}
{Ho}, W. C.~G. 2007, \mnras, 380, 71

\bibitem[{{Ho}(2014)}]{Ho14}
{Ho}, W. C.~G. 2014, in Magnetic Fields throughout Stellar Evolution, ed.
  P.~{Petit}, M.~{Jardine}, \& H.~C. {Spruit}, Vol. 302, 435--438

\bibitem[{{Ho} {et~al.}(2015){Ho}, {Elshamouty}, {Heinke}, \&
  {Potekhin}}]{Hoetal15}
{Ho}, W. C.~G., {Elshamouty}, K.~G., {Heinke}, C.~O., \& {Potekhin}, A.~Y.
  2015, \prc, 91, 015806

\bibitem[{{Ho} {et~al.}(2012{\natexlab{a}}){Ho}, {Glampedakis}, \&
  {Andersson}}]{Hoetal12b}
{Ho}, W. C.~G., {Glampedakis}, K., \& {Andersson}, N. 2012{\natexlab{a}},
  \mnras, 425, 1600

\bibitem[{{Ho} {et~al.}(2012{\natexlab{b}}){Ho}, {Glampedakis}, \&
  {Andersson}}]{Hoetal12}
---. 2012{\natexlab{b}}, \mnras, 422, 2632

\bibitem[{{Ho} \& {Heinke}(2009)}]{HoHeinke09}
{Ho}, W. C.~G., \& {Heinke}, C.~O. 2009, \nat, 462, 71

\bibitem[{{Ho} {et~al.}(2007){Ho}, {Kaplan}, {Chang}, {van Adelsberg}, \&
  {Potekhin}}]{Hoetal07}
{Ho}, W. C.~G., {Kaplan}, D.~L., {Chang}, P., {van Adelsberg}, M., \&
  {Potekhin}, A.~Y. 2007, \mnras, 375, 821

\bibitem[{{Ho} {et~al.}(2008){Ho}, {Potekhin}, \& {Chabrier}}]{Hoetal08}
{Ho}, W. C.~G., {Potekhin}, A.~Y., \& {Chabrier}, G. 2008, \apjs, 178, 102

\bibitem[{{Hobbs} {et~al.}(2020){Hobbs}, {Manchester}, {Dunning}, {Jameson},
  {Roberts}, {George}, {Green}, {Tuthill}, {Toomey}, {Kaczmarek}, {Mader},
  {Marquarding}, {Ahmed}, {Amy}, {Bailes}, {Beresford}, {Bhat}, {Bock},
  {Bourne}, {Bowen}, {Brothers}, {Cameron}, {Carretti}, {Carter}, {Castillo},
  {Chekkala}, {Cheng}, {Chung}, {Craig}, {Dai}, {Dawson}, {Dempsey}, {Doherty},
  {Dong}, {Edwards}, {Ergesh}, {Gao}, {Han}, {Hayman}, {Indermuehle},
  {Jeganathan}, {Johnston}, {Kanoniuk}, {Kesteven}, {Kramer}, {Leach},
  {Mcintyre}, {Moss}, {Os{\l}owski}, {Phillips}, {Pope}, {Preisig}, {Price},
  {Reeves}, {Reilly}, {Reynolds}, {Robishaw}, {Roush}, {Ruckley}, {Sadler},
  {Sarkissian}, {Severs}, {Shannon}, {Smart}, {Smith}, {Smith}, {Sobey},
  {Staveley-Smith}, {Tzioumis}, {van Straten}, {Wang}, {Wen}, \&
  {Whiting}}]{UWL}
{Hobbs}, G., {Manchester}, R.~N., {Dunning}, A., {et~al.} 2020, \pasa, 37, e012

\bibitem[{{Hurley-Walker} {et~al.}(2017){Hurley-Walker}, {Callingham},
  {Hancock}, {Franzen}, {Hindson}, {Kapi{\'n}ska}, {Morgan}, {Offringa},
  {Wayth}, {Wu}, {Zheng}, {Murphy}, {Bell}, {Dwarakanath}, {For}, {Gaensler},
  {Johnston-Hollitt}, {Lenc}, {Procopio}, {Staveley-Smith}, {Ekers}, {Bowman},
  {Briggs}, {Cappallo}, {Deshpande}, {Greenhill}, {Hazelton}, {Kaplan},
  {Lonsdale}, {McWhirter}, {Mitchell}, {Morales}, {Morgan}, {Oberoi}, {Ord},
  {Prabu}, {Shankar}, {Srivani}, {Subrahmanyan}, {Tingay}, {Webster},
  {Williams}, \& {Williams}}]{Hurleywalkeretal17}
{Hurley-Walker}, N., {Callingham}, J.~R., {Hancock}, P.~J., {et~al.} 2017,
  \mnras, 464, 1146

\bibitem[{{Igoshev}(2020)}]{Igoshev2020}
{Igoshev}, A.~P. 2020, \mnras, 494, 3663

\bibitem[{{Kaminker} {et~al.}(2006){Kaminker}, {Yakovlev}, {Potekhin},
  {Shibazaki}, {Shternin}, \& {Gnedin}}]{Kaminkeretal06}
{Kaminker}, A.~D., {Yakovlev}, D.~G., {Potekhin}, A.~Y., {et~al.} 2006, \mnras,
  371, 477

\bibitem[{{Kantor} \& {Gusakov}(2007)}]{KantorGusakov07}
{Kantor}, E.~M., \& {Gusakov}, M.~E. 2007, \mnras, 381, 1702

\bibitem[{{Lattimer} {et~al.}(1991){Lattimer}, {Pethick}, {Prakash}, \&
  {Haensel}}]{Lattimeretal91}
{Lattimer}, J.~M., {Pethick}, C.~J., {Prakash}, M., \& {Haensel}, P. 1991,
  \prl, 66, 2701

\bibitem[{{Leinson}(2010)}]{Leinson10}
{Leinson}, L.~B. 2010, \prc, 81, 025501

\bibitem[{{Luo} {et~al.}(2019){Luo}, {Ransom}, {Demorest}, {van Haasteren},
  {Ray}, {Stovall}, {Bachetti}, {Archibald}, {Kerr}, {Colen}, \&
  {Jenet}}]{pint_software}
{Luo}, J., {Ransom}, S., {Demorest}, P., {et~al.} 2019

\bibitem[{{Luo} {et~al.}(2021){Luo}, {Ransom}, {Demorest}, {Ray}, {Archibald},
  {Kerr}, {Jennings}, {Bachetti}, {van Haasteren}, {Champagne}, {Colen},
  {Phillips}, {Zimmerman}, {Stovall}, {Lam}, \& {Jenet}}]{Luoetal21}
---. 2021, \apj, 911, 45

\bibitem[{{Mart{\'\i}n} {et~al.}(2016){Mart{\'\i}n}, {Torres}, \&
  {Pedaletti}}]{Martinetal16}
{Mart{\'\i}n}, J., {Torres}, D.~F., \& {Pedaletti}, G. 2016, \mnras, 459, 3868

\bibitem[{{Mayer} \& {Becker}(2024)}]{MayerBecker24}
{Mayer}, M. G.~F., \& {Becker}, W. 2024, \aap, 684, A208

\bibitem[{{McGowan} {et~al.}(2006){McGowan}, {Zane}, {Cropper}, {Vestrand}, \&
  {Ho}}]{Mcgowanetal06}
{McGowan}, K.~E., {Zane}, S., {Cropper}, M., {Vestrand}, W.~T., \& {Ho}, C.
  2006, \apj, 639, 377

\bibitem[{{McLaughlin} {et~al.}(2002){McLaughlin}, {Arzoumanian}, {Cordes},
  {Backer}, {Lommen}, {Lorimer}, \& {Zepka}}]{j1740_mam}
{McLaughlin}, M.~A., {Arzoumanian}, Z., {Cordes}, J.~M., {et~al.} 2002, \apj,
  564, 333

\bibitem[{{Morello} {et~al.}(2019){Morello}, {Barr}, {Cooper}, {Bailes},
  {Bates}, {Bhat}, {Burgay}, {Burke-Spolaor}, {Cameron}, {Champion}, {Eatough},
  {Flynn}, {Jameson}, {Johnston}, {Keith}, {Keane}, {Kramer}, {Levin}, {Ng},
  {Petroff}, {Possenti}, {Stappers}, {van Straten}, \& {Tiburzi}}]{CLFD}
{Morello}, V., {Barr}, E.~D., {Cooper}, S., {et~al.} 2019, \mnras, 483, 3673

\bibitem[{{Mori} \& {Ho}(2007)}]{MoriHo07}
{Mori}, K., \& {Ho}, W. C.~G. 2007, \mnras, 377, 905

\bibitem[{{Motch} {et~al.}(2003){Motch}, {Zavlin}, \& {Haberl}}]{Motchetal03}
{Motch}, C., {Zavlin}, V.~E., \& {Haberl}, F. 2003, \aap, 408, 323

\bibitem[{{Noll}(2010)}]{igsg_2}
{Noll}, C.~E. 2010, Advances in Space Research, 45, 1421

\bibitem[{{Page} \& {Applegate}(1992)}]{PageApplegate92}
{Page}, D., \& {Applegate}, J.~H. 1992, \apjl, 394, L17

\bibitem[{{Page} {et~al.}(2004){Page}, {Lattimer}, {Prakash}, \&
  {Steiner}}]{Pageetal04}
{Page}, D., {Lattimer}, J.~M., {Prakash}, M., \& {Steiner}, A.~W. 2004, \apjs,
  155, 623

\bibitem[{{Pavlov} {et~al.}(1995){Pavlov}, {Shibanov}, {Zavlin}, \&
  {Meyer}}]{Pavlovetal95}
{Pavlov}, G.~G., {Shibanov}, Y.~A., {Zavlin}, V.~E., \& {Meyer}, R.~D. 1995, in
  NATO Advanced Study Institute (ASI) Series C, Vol. 450, The Lives of the
  Neutron Stars, ed. M.~A. {Alpar}, U.~{Kiziloglu}, \& J.~{van Paradijs}, 71

\bibitem[{{Pearson} {et~al.}(2018){Pearson}, {Chamel}, {Potekhin}, {Fantina},
  {Ducoin}, {Dutta}, \& {Goriely}}]{Pearsonetal18}
{Pearson}, J.~M., {Chamel}, N., {Potekhin}, A.~Y., {et~al.} 2018, \mnras, 481,
  2994

\bibitem[{{Pol} {et~al.}(2021){Pol}, {Burke-Spolaor}, {Hurley-Walker},
  {Blumer}, {Johnston}, {Keith}, {Keane}, {Burgay}, {Possenti}, {Petroff}, \&
  {Bhat}}]{Poletal21}
{Pol}, N., {Burke-Spolaor}, S., {Hurley-Walker}, N., {et~al.} 2021, \apj, 911,
  121

\bibitem[{{Potekhin}(2014)}]{Potekhin14}
{Potekhin}, A.~Y. 2014, Physics Uspekhi, 57, 735

\bibitem[{{Potekhin} \& {Chabrier}(2018)}]{PotekhinChabrier18}
{Potekhin}, A.~Y., \& {Chabrier}, G. 2018, \aap, 609, A74

\bibitem[{{Potekhin} {et~al.}(2015){Potekhin}, {Pons}, \&
  {Page}}]{Potekhinetal15}
{Potekhin}, A.~Y., {Pons}, J.~A., \& {Page}, D. 2015, \ssr, 191, 239

\bibitem[{{Potekhin} {et~al.}(2012){Potekhin}, {Suleimanov}, {van Adelsberg},
  \& {Werner}}]{Potekhinetal12}
{Potekhin}, A.~Y., {Suleimanov}, V.~F., {van Adelsberg}, M., \& {Werner}, K.
  2012, \aap, 546, A121

\bibitem[{{Potekhin} {et~al.}(2020){Potekhin}, {Zyuzin}, {Yakovlev},
  {Beznogov}, \& {Shibanov}}]{Potekhinetal20}
{Potekhin}, A.~Y., {Zyuzin}, D.~A., {Yakovlev}, D.~G., {Beznogov}, M.~V., \&
  {Shibanov}, Y.~A. 2020, \mnras, 496, 5052

\bibitem[{{Ransom} {et~al.}(2002){Ransom}, {Eikenberry}, \&
  {Middleditch}}]{Ransometal02}
{Ransom}, S.~M., {Eikenberry}, S.~S., \& {Middleditch}, J. 2002, \aj, 124, 1788

\bibitem[{{Schwenk} {et~al.}(2003){Schwenk}, {Friman}, \&
  {Brown}}]{Schwenketal03}
{Schwenk}, A., {Friman}, B., \& {Brown}, G.~E. 2003, \nphysa, 713, 191

\bibitem[{{Shternin} {et~al.}(2019){Shternin}, {Kirichenko}, {Zyuzin}, {Yu},
  {Danilenko}, {Voronkov}, \& {Shibanov}}]{Shterninetal19}
{Shternin}, P., {Kirichenko}, A., {Zyuzin}, D., {et~al.} 2019, \apj, 877, 78

\bibitem[{{Shternin} {et~al.}(2018){Shternin}, {Baldo}, \&
  {Haensel}}]{Shterninetal18}
{Shternin}, P.~S., {Baldo}, M., \& {Haensel}, P. 2018, Physics Letters B, 786,
  28

\bibitem[{{Skiathas} \& {Gourgouliatos}(2024)}]{SkiathasGourgouliatos24}
{Skiathas}, D., \& {Gourgouliatos}, K.~N. 2024, \mnras, 528, 5178

\bibitem[{{Sotomayor-Beltran} {et~al.}(2013){Sotomayor-Beltran}, {Sobey},
  {Hessels}, {de Bruyn}, {Noutsos}, {Alexov}, {Anderson}, {Asgekar}, {Avruch},
  {Beck}, {Bell}, {Bell}, {Bentum}, {Bernardi}, {Best}, {Birzan}, {Bonafede},
  {Breitling}, {Broderick}, {Brouw}, {Br{\"u}ggen}, {Ciardi}, {de Gasperin},
  {Dettmar}, {van Duin}, {Duscha}, {Eisl{\"o}ffel}, {Falcke}, {Fallows},
  {Fender}, {Ferrari}, {Frieswijk}, {Garrett}, {Grie{\ss}meier}, {Grit},
  {Gunst}, {Hassall}, {Heald}, {Hoeft}, {Horneffer}, {Iacobelli}, {Juette},
  {Karastergiou}, {Keane}, {Kohler}, {Kramer}, {Kondratiev}, {Koopmans},
  {Kuniyoshi}, {Kuper}, {van Leeuwen}, {Maat}, {Macario}, {Markoff}, {McKean},
  {Mulcahy}, {Munk}, {Orru}, {Paas}, {Pandey-Pommier}, {Pilia}, {Pizzo},
  {Polatidis}, {Reich}, {R{\"o}ttgering}, {Serylak}, {Sluman}, {Stappers},
  {Tagger}, {Tang}, {Tasse}, {ter Veen}, {Vermeulen}, {van Weeren}, {Wijers},
  {Wijnholds}, {Wise}, {Wucknitz}, {Yatawatta}, \& {Zarka}}]{ionFR}
{Sotomayor-Beltran}, C., {Sobey}, C., {Hessels}, J.~W.~T., {et~al.} 2013, \aap,
  552, A58

\bibitem[{Susobhanan {et~al.}(2024)Susobhanan, Kaplan, Archibald, Luo, Ray,
  Pennucci, Ransom, Agazie, Fiore, Larsen, O'Neill, van Haasteren,
  Anumarlapudi, Bachetti, Bhakta, Champagne, Cromartie, Demorest, Jennings,
  Kerr, Levina, McEwen, Shapiro-Albert, \& Swiggum}]{pint_paper_2}
Susobhanan, A., Kaplan, D., Archibald, A., {et~al.} 2024, arXiv:2405.01977

\bibitem[{{Takatsuka} \& {Tamagaki}(2004)}]{TakatsukaTamagaki04}
{Takatsuka}, T., \& {Tamagaki}, R. 2004, Progress of Theoretical Physics, 112,
  37

\bibitem[{{Th{\'e}bault} {et~al.}(2015){Th{\'e}bault}, {Finlay}, {Beggan},
  {Alken}, {Aubert}, {Barrois}, {Bertrand}, {Bondar}, {Boness}, {Brocco},
  {Canet}, {Chambodut}, {Chulliat}, {Co{\"\i}sson}, {Civet}, {Du}, {Fournier},
  {Fratter}, {Gillet}, {Hamilton}, {Hamoudi}, {Hulot}, {Jager}, {Korte},
  {Kuang}, {Lalanne}, {Langlais}, {L{\'e}ger}, {Lesur}, {Lowes}, {Macmillan},
  {Mandea}, {Manoj}, {Maus}, {Olsen}, {Petrov}, {Ridley}, {Rother}, {Sabaka},
  {Saturnino}, {Schachtschneider}, {Sirol}, {Tangborn}, {Thomson},
  {T{\o}ffner-Clausen}, {Vigneron}, {Wardinski}, \& {Zvereva}}]{igrf_1}
{Th{\'e}bault}, E., {Finlay}, C.~C., {Beggan}, C.~D., {et~al.} 2015, Earth,
  Planets and Space, 67, 79

\bibitem[{{van Adelsberg} {et~al.}(2005){van Adelsberg}, {Lai}, {Potekhin}, \&
  {Arras}}]{vanAdelsbergetal05}
{van Adelsberg}, M., {Lai}, D., {Potekhin}, A.~Y., \& {Arras}, P. 2005, \apj,
  628, 902

\bibitem[{{Verner} {et~al.}(1996){Verner}, {Ferland}, {Korista}, \&
  {Yakovlev}}]{Verneretal96}
{Verner}, D.~A., {Ferland}, G.~J., {Korista}, K.~T., \& {Yakovlev}, D.~G. 1996,
  \apj, 465, 487

\bibitem[{{Vigan{\`o}} {et~al.}(2013){Vigan{\`o}}, {Rea}, {Pons}, {Perna},
  {Aguilera}, \& {Miralles}}]{Viganoetal13}
{Vigan{\`o}}, D., {Rea}, N., {Pons}, J.~A., {et~al.} 2013, \mnras, 434, 123

\bibitem[{{Wilms} {et~al.}(2000){Wilms}, {Allen}, \& {McCray}}]{Wilmsetal00}
{Wilms}, J., {Allen}, A., \& {McCray}, R. 2000, \apj, 542, 914

\bibitem[{{Yao} {et~al.}(2017){Yao}, {Manchester}, \& {Wang}}]{Yaoetal17}
{Yao}, J.~M., {Manchester}, R.~N., \& {Wang}, N. 2017, \apj, 835, 29

\bibitem[{{Yar-Uyaniker} {et~al.}(2004){Yar-Uyaniker}, {Uyaniker}, \&
  {Kothes}}]{Yaruyanikeretal04}
{Yar-Uyaniker}, A., {Uyaniker}, B., \& {Kothes}, R. 2004, \apj, 616, 247

\end{thebibliography}

\end{document}